\documentclass[11pt]{article}
\usepackage[margin=3cm]{geometry}

\usepackage{amssymb}

\usepackage{amsmath}
\usepackage{url}
\usepackage{color}

\usepackage{comment}

\usepackage{pdfsync}

\newcommand{\netpar}{\mid\mid}
\newcommand{\val}[1]{\langle\,#1\,\rangle}

\newcommand{\ppar}{\mid}
\newcommand{\Let}{{\bf let}}
\newcommand{\Letin}{{\bf in}}

\newcommand{\Out}[2]{{\bf out}({#1})@{\it #2}}
\newcommand{\In}[2]{{\bf in}({#1})@{\it #2}}
\newcommand{\Read}[2]{{\bf read}({#1})@{\it #2}}
\newcommand{\Eval}[2]{{\bf eval}({#1})@{\it #2}}

\newcommand{\nil}{{\bf 0}}

\newcommand{\Test}[2]{{\bf test}({#1})@{\it #2}}

\newcommand{\Proceed}{{\bf proceed}}

\newcommand{\adefn}[3]{#1[#2]\triangleq#3}
\newcommand{\YZl}{l}
\newcommand{\Fail}{{\sf fail}}

\newcommand{\VF}[1]{{\it cl}($#1$)}

\newcommand{\Lnt}{\ell^\lambda}
\newcommand{\Ln}{\ell}

\newcommand{\Lnc}{\ell_t}
\newcommand{\Lntc}{\ell_t^\lambda}

\newcommand{\Lc}{\ell}

\newcommand{\Lb}{\ell}
\newcommand{\Lat}{\ell^\lambda}

\newcommand{\veck}[1]{\overrightarrow{#1}}
\newcommand{\Break}{{\bf break}}

\newcommand{\Case}{{\bf case}}
\newcommand{\TT}{{\bf tt}}
\newcommand{\FF}{{\bf ff}}

\newcommand{\doublebracketleft} {[\![}
\newcommand{\doublebracketright}{]\!]}
\newcommand{\db}[1]{\doublebracketleft #1 \doublebracketright}

\newcommand{\md}{MgDavis}
\newcommand{\ds}{DrSmith}
\newcommand{\no}{NsOlsen}
\newcommand{\dn}{DrHansen}
\newcommand{\dsn}{DrJensen}
\newcommand{\rmi}{RsMiller}
\newcommand{\al}{Alice}

\newcommand{\un}{\underline{ \ \ }}

\newcommand{\MR}{MedicalRecord}
\newcommand{\PN}{PrivateNote}
\newcommand{\RE}{Recent}
\newcommand{\PA}{Past}
\newcommand{\EDB}{EHDB}
\newcommand{\EDBT}{EHDB2}
\newcommand{\ER}{RDB}
\newcommand{\EP}{PDB}
\newcommand{\ECL}{CLDB}
\newcommand{\EAL}{ALDB}

\newcommand{\changefont}{\small}

\newcommand{\Inference}[2]{\begin{array}{@{}c@{}}#1\\[0em]\hline\\[-0.9em]#2\\
\end{array}}

\newenvironment{ARRAY}[1]{%
  \begin{tabular*}{\textwidth}{@{\extracolsep{\fill}}c@{}c@{}c}
  \hline
  &&\\[-5pt]
  &\begin{math}\begin{array}{@{} #1 @{}}
}
{ \end{array}\end{math}&\\
  &&\\[-5pt]
  \hline
  \end{tabular*}
}

\newtheorem{examples}{Example}

\newenvironment{qquote}{\vspace*{-2ex}\begin{quote}}{\end{quote}\vspace*{-2ex}}
\begin{document}




\title{Secondary use of data in EHR systems} 


\author{Fan Yang\footnote{DTU Informatics, Technical University of Denmark, Lyngby, Denmark}
\and Chris Hankin\footnote{ISST, Imperial College London, London United Kingdom}
\and  Flemming Nielson \footnote{DTU Informatics, Technical University of Denmark, Lyngby, Denmark}
\and Hanne Riis Nielson\footnote{DTU Informatics, Technical University of Denmark, Lyngby, Denmark}}











\maketitle

\begin{abstract}
We show how to use aspect-oriented programming to separate security and trust 
issues from the logical design of mobile, distributed systems.  The main challenge is how to enforce various types of security policies, in particular predictive access control policies - policies based on the future behavior of a program. A novel feature of our approach is that
advice is able to analyze the future use of data.  We consider a number of different security policies, concerning both primary and secondary use of data, some of which can only be enforced by
analysis of process continuations.
\end{abstract}

\section{Introduction}\label{sec:introduction}

Whilst there is broad agreement that security and other non-functional
properties should be designed into systems {\it ab initio} it is also recognized that, as society becomes more IT-savvy, our expectations about security and privacy evolve.  This is usually followed by changes in regulation in the form of standards and legislation. Thus, whilst we would still argue that security should feature in the initial design of a system, there is merit in separating out security and other non-functional properties so that they can be updated without disturbing the functional aspects of the system. 

This paper focuses on designing a language for specifying policies for access control and explicit flow of information. The traditional approach to enforcing such security policies is to use a \emph{reference monitor} \cite{Gollmann199902} that dynamically
tracks the execution of the program; it makes appropriate checks on
each basic operation being performed, either blocking the operation or
allowing it to proceed. In concrete systems this is implemented as
part of the operating system or as part of the interpreter for the
language at hand (e.g. the Java byte code interpreter); in both cases
as part of the trusted computing base. Sometimes it is found to be more cost effective to systematically modify the code so as to explicitly perform the checks that the reference monitor would otherwise have imposed \cite{erlingsson2000iej}. In any case, even small modification in the security policies may involve substantial changes in the code or the underlying system.

The notion of \emph{aspect-oriented programming} \cite{kiczales2001oa,kiczales97aspectoriented} is an interesting approach to separation of concerns. The enforcement of security policies is an obvious candidate for such separation of concerns, e.g.~because the security policy can be implemented by more skilled or more trusted programmers, or indeed
because security considerations can be retrofitted by (re)defining
\emph{advice} to suit the (new) security policy.  The detailed definition of the advice will then make decisions about how to possibly modify the operation being trapped. This calls for a modified language (like AspectJ \cite{kiczales2001oa} for Java) that supports the use of aspects and where a notion of trapping operations and applying advice has been incorporated. It is possible to systematically modify the code so as to explicitly perform the operations that the advice would have imposed (e.g.~\cite{kiczales2001oa}).

In many cases the aspect-oriented approach provides a more flexible
way for dealing with modifications in security policies
\cite{Daniel07,gao2004aao,georg2002uad,phung:spe,dewin2004dsa}
than the use of reference monitors. 
It facilitates the use of frameworks for security policies that may be well suited to the task at hand but that are perhaps not of general applicability and therefore not appropriate for incorporating into a reference monitor.

\paragraph{Outline of the paper.}

In this paper we are primarily interested in the modelling of mobile, distributed systems. The work is based on the coordination language KLAIM \cite{De1998} (reviewed in Section \ref{sec:klaim}) that facilitates distribution of data, mobility of code and handling of dynamically evolving, open systems. The main contribution of the paper is the design of AspectKE, an aspect-oriented extension of KLAIM that facilitates the trapping of actions (presented in Section \ref{sec:review}) as well as processes (presented in Section \ref{sec:process}), which can enforce traditional access control policies (presented in Section \ref{sec:access}) as well as  predictive access control policies (presented in Section \ref{sec:direct}) -- i.e.,  security policies based on the future behavior of a program. 

To evaluate our language design we shall throughout the paper illustrate its features using a running example based on a health information system for a care facility for the elderly people in New South Wales, Australia \cite{evered2004csa}. In Section \ref{sec:access} we show how to use AspectKE to enforce basic \emph{primary use of data policies}, that is, policies concerned with the right to access data. Here we consider the three classical access control models, namely discretionary access control, mandatory access control and role-based access control \cite{Gollmann199902}. Furthermore, we illustrate how multiple security policies can be integrated into existing systems thereby allowing policies to be refined at later stages in the system development. 
In Section \ref{sec:direct} we show how \emph{secondary use of data} policies can be modelled; these policies are concerned with how data is used once it has been obtained \cite{safran2007tnf}. They can be enforced by using predictive access control policy enforcement mechanism offered by AspectKE. Here we exploit the ability to analyze not only the behavior of remotely executed processes but also the future use of data. To the best of our knowledge few, if any,
proposals have ever used aspect oriented programming to tackle secondary use of data policies and provide a predictive access control policy enforcement mechanism. 

Finally, in Section \ref{sec:related} we present related work and we conclude in Section \ref{sec:conclusion} with a discussion of the experience gathered from a proof-of-concept implementation \cite{fan10} and outline some future work.

\section{Background: KLAIM} \label{sec:klaim}

AspectKE is an extension of the KLAIM (\emph{Kernel Language for Agents Interaction and Mobility}) coordination language \cite{De1998} with  support for aspect oriented programming. In this section we will review the fragment of KLAIM that will be used for AspectKE in the following section.   

KLAIM is a language specifically designed to program distributed systems consisting of several mobile components that interact through multiple distributed tuple spaces (databases). KLAIM uses a Linda-like generative communication model \cite{David85} but, instead of using Linda's global shared tuple space (shared database), KLAIM associates a local tuple space with each node of a net. Each node may also have  processes associated with it; the KLAIM computing primitives allow programmers to distribute and retrieve data and processes to and from locations (nodes) of a net, evaluate processes at remote locations and introduce new locations to the net.


\subsection{Syntax of KLAIM}

The syntax of a fragment of KLAIM is displayed in Table \ref{klaim:syn_NetProcess}. 


\begin{table}[t]
	\changefont
$$
\begin{ARRAY}{l@{\qquad}rcl}
N  \in  {\bf Net} &
N & ::= & N_1\netpar N_2\ \mid\ \YZl::P\ \mid\
\YZl::\val{\veck{\YZl}}
\\[1ex]
P  \in  {\bf Proc} &
P & ::= &  P_1\ppar P_2\ \mid \sum_i a_i.P_i\ \mid\ *P
\\[1ex]
a  \in  {\bf Act} &
a & ::= & \Out{\veck{\ell}}{\ell} \mid \In{\veck{\ell^{\lambda}}}{\ell} \mid
   \Read{\veck{\ell^\lambda}}{\ell} \mid \Eval{P}{\ell} \mid \mathbf{newloc}(!u)
\\[1ex]
c  \in  {\bf Cap} &
c & ::= & \mathbf{out} \mid \mathbf{in} \mid \mathbf{read} \mid \mathbf{eval} \mid \mathbf{newloc}
\\[1ex]
\Ln,\Lnt \ \in \ {\bf Loc} &
\Ln & ::= & u \  \mid\ \YZl \hfill 
  \Lnt\  ::= \  \Ln \mid\ !u  
\end{ARRAY}
$$
\caption{KLAIM Syntax -- Nets, Processes and Actions (Part of AspectKE).} \label{klaim:syn_NetProcess}
\end{table}

A net (in {\bf Net}) is a parallel composition of located processes and/or located tuples. For simplicity, components of tuples can be location constants only\footnote{Compared with the original KLAIM, we do not allow processes to be components of tuples.}. We use the notation 
$\veck{\YZl}$ to represent a sequence of location constants and $\epsilon$ is used to represent the empty sequence. Nets must be \emph{closed}: all variables must be in the  scope of a defining occurrence (indicated by an exclamation mark). 

A process (in {\bf Proc}) can be a parallel composition of processes, a guarded sum of  action prefixed processes, or a replicated process  (indicated by the $*$ operator). We write \textbf{0} for a nullary sum, $a.P$ for a unary sum, and $a_1.P_1 + a_2.P_2 $ for a binary sum. 

An action (in {\bf Act}) operates on locations, tuples and processes: a tuple can be output to, input from  (read and delete the source) and read from (read and keep the source)  a location;  processes can be spawned at a location;  new locations can also be created.  The actual operation performed by an action is called a \emph{capability} (in {\bf Cap}) -- this is a key concept when formalizing uses of data later. We do not distinguish real locations and data: all  of them are called locations (in {\bf Loc}) in our setting, which can be location  constants $l$, defining (i.e. binding) occurrences of location variables $!u$  (where the scope is the entire process to the right of the occurrence),  and use of location variables $u$.


\paragraph{Well-Formedness of Locations and Actions.}

We do not allow multiple defining
occurrences of the same variable in an action. We also prohibit
bound variables and free variables from sharing any name
in a single action.  Thus we disallow  ${\bf in}(!u,!u)@l$ as well as   ${\bf
in}(!u,u)@l$.



\subsection{Semantics of KLAIM}

Informally the meaning of a KLAIM program is as follows:
\begin{enumerate}
\item\label{start} a node is selected for the next step of execution
\item if the process at the node is a choice, then one of the
enabled choices is chosen non-deterministically and executed
as described in the following four steps
\item if the prefix of the process is an output action, the output is
performed
\item if the prefix of the process is an input (either destructive
or non-destructive), the input action is enabled if there is a matching tuple at 
the target location,
and the input is performed and appropriate variables are bound in
the remainder of the process
\item if the prefix is an eval, the process is spawned at the target location
\item if the prefix is a newloc, the network is dynamically extended
with a new location and the continuation process is given the address of that
location
\item then return to Step \ref{start}
\end{enumerate}
Notice that we do not need to deal with parallelism and
replication within nodes because,
at the cost of having duplicate addresses in the network, these can be 
lifted to
the net level.

\subsection{Running Example}\label{runningexample}

Health Care Information Systems are gradually becoming prevalent and indispensable to our society. An \emph{electronic health record} (EHR), part of a system's database, stores a patient's data and is created, developed, and maintained by the health care providers.

To illustrate the use of KLAIM, we now introduce a typical EHR system, which is inspired by \cite{evered2004csa}, and the scenario presented here is used throughout the paper. 

The EHR database ({\sf \EDB}) stores all patient healthcare records and we assume that there are two types of data recorded for each patient: medical records ({\sf MedicalRecord}) and private notes ({\sf PrivateNote}).  Medical records are entries created by doctors and so are the private notes; however the latter are of a more confidential nature. Also we distinguish between past records ({\sf Past}) that have been entered into the EHR system previously and recent records ({\sf Recent}) that have been created since the patient was admitted to the hospital. We therefore assume that the EHR database contains tuples with the following five fields: 
$$\small
\begin{tabular}{|l|l|}
\hline
\textit{patient} & The name of the patient\\
\hline
\textit{recordtype} & The type of record: {\sf \MR} or {\sf \PN}\\
\hline
\textit{author} & The author of the record \\
\hline
\textit{createdtime} & The time of creation of the record: {\sf \RE} or {\sf \PA}\\
\hline
\textit{subject} & The record's content\\
\hline
\end{tabular}
$$
For example $\langle{\sf \al, \MR, \ds, \RE, text}\rangle$  is a recent medical record of ${\sf \al}$,  created by ${\sf \ds}$ and it has content ${\sf text}$. 

Doctors and nurses, as well as the patient, can access a patient's record. We
 model these actors as locations in a network; the process at the location
represents the actions of the individual and the data is the individual's
local ``knowledge''.   As an example the following process expresses that {\sf \ds} reads one of the {\sf \PA} medical records for {\sf \al} created by {\sf \dn} before she was admitted to this hospital, writes some of the information in her own note (in location {\sf \ds}) and then creates a new medical record for the patient: 
$$\small
\begin{array}{ll}
{\sf \ds} :: & {\bf read}({\sf Alice,\MR, \dn,\PA},!content)@{\sf \EDB}.\\ 
& {\bf out}({\sf \al},content)@{\sf \ds}.\\
&{\bf out}({\sf \al,\MR, \ds ,\RE,newtext})@{\sf \EDB}
\end{array}
$$ 
Here {\sf \ds} will first consult location {\sf \EDB} and read a five-tuple whose first four components are {\sf \al,\MR,\dn}, and {\sf \PA} respectively and the corresponding fifth component is assigned to variable $content$. The second action will write the $content$ read at the first action to the location associated with {\sf \ds}. The final construct will write a new five-tuple to location {\sf \EDB} for this patient whose last three components indicate that the author is {\sf \ds}, it is a {\sf \RE} medical record and the content is {\sf newtext}.

To illustrate the semantics of KLAIM let us consider the following net, consisting of locations {\sf \EDB} and {\sf \ds}:
$$\small
\begin{array}{ll}
& {\sf \EDB} :: \langle {\sf Alice}, {\sf MedicalRecord}, {\sf \dn}, {\sf \PA}, {\sf alicetext} \rangle  \\
\netpar
& {\sf \EDB} :: \langle {\sf Bob}, {\sf PrivateNote}, {\sf \dsn}, {\sf \RE}, {\sf bobtext} \rangle  \\
\netpar
& {\sf \ds} :: \ {\bf read}({\sf Alice}, {\sf MedicalRecord}, {\sf \dn}, {\sf \PA},!content)@{\sf \EDB}.\\
& \qquad \qquad  \ {\bf out}({\sf Alice},content)@{\sf \ds}.\\ 
& \qquad \qquad  \ {\bf out}({\sf Alice}, {\sf MedicalRecord}, {\sf \ds}, {\sf \RE},{\sf newtext})@{\sf \EDB}
\end{array}
$$
The execution may proceed as follows:
$$\small
\begin{array}{ll}
& {\sf \EDB} :: \langle {\sf Alice}, {\sf MedicalRecord}, {\sf \dn}, {\sf \PA}, {\sf alicetext} \rangle  \\
\netpar
& {\sf \EDB} :: \langle {\sf Bob}, {\sf PrivateNote}, {\sf \dsn}, {\sf \RE}, {\sf bobtext} \rangle  \\
\netpar
& {\sf \ds} :: \ {\bf read}({\sf Alice}, {\sf MedicalRecord}, {\sf \dn}, {\sf \PA},!content)@{\sf \EDB}.\\
& \qquad \qquad  \ {\bf out}({\sf Alice},content)@{\sf \ds}.\\ 
& \qquad \qquad  \ {\bf out}({\sf Alice}, {\sf MedicalRecord}, {\sf \ds}, {\sf \RE},{\sf newtext})@{\sf \EDB}\\

\rightarrow\\
& {\sf \EDB} :: \langle {\sf Alice}, {\sf MedicalRecord}, {\sf \dn}, {\sf \PA}, {\sf alicetext} \rangle  \\
\netpar
& {\sf \EDB} :: \langle {\sf Bob}, {\sf PrivateNote}, {\sf \dsn}, {\sf \RE}, {\sf bobtext} \rangle  \\
\netpar
& {\sf \ds} :: \ {\bf out}({\sf Alice},{\sf alicetext})@{\sf \ds}.\\ 
& \qquad \qquad  \ {\bf out}({\sf Alice}, {\sf MedicalRecord}, {\sf \ds}, {\sf \RE},{\sf newtext})@{\sf \EDB}
\\

\rightarrow\\
& {\sf \EDB} :: \langle {\sf Alice}, {\sf MedicalRecord}, {\sf \dn}, {\sf \PA}, {\sf alicetext} \rangle  \\
\netpar
& {\sf \EDB} :: \langle {\sf Bob}, {\sf PrivateNote}, {\sf \dsn}, {\sf \RE}, {\sf bobtext} \rangle  \\
\netpar
& {\sf \ds} :: \langle {\sf Alice}, {\sf alicetext} \rangle  \\
\netpar
& {\sf \ds} :: \  {\bf out}({\sf Alice}, {\sf MedicalRecord}, {\sf \ds}, {\sf \RE},{\sf newtext})@{\sf \EDB}

\end{array}
$$

$$\small
\begin{array}{ll}

\rightarrow\\
& {\sf \EDB} :: \langle {\sf Alice}, {\sf MedicalRecord}, {\sf \dn}, {\sf \PA}, {\sf alicetext} \rangle  \\
\netpar
& {\sf \EDB} :: \langle {\sf Alice}, {\sf MedicalRecord}, {\sf \ds}, {\sf \RE}, {\sf newtext} \rangle  \\
\netpar
& {\sf \EDB} :: \langle {\sf Bob}, {\sf PrivateNote}, {\sf \dsn}, {\sf \RE}, {\sf bobtext} \rangle  \\
\netpar
& {\sf \ds} :: \langle {\sf Alice}, {\sf alicetext} \rangle  \\

\end{array}
$$
{\sf \ds} first reads the tuple
$\langle {\sf Alice}, {\sf MedicalRecord}, {\sf \dn}, {\sf \PA}, {\sf alicetext} \rangle$ from {\sf \EDB}; the binding of the
variable $content$ is reflected in the continuation of the
process. In the second step {\sf \ds} outputs a tuple that consists of {\sf Alice} together with a bound $content$ ({\sf alicetext})  to her own tuple space. In the final step, a new tuple that represents a new medical record is written to location {\sf \EDB}.


\section{AspectKE: Trapping Actions} \label{sec:review}

We now show how to integrate aspects into KLAIM by presenting the basic features of AspectKE, with a focus on how aspects trap actions 
in a KLAIM program. 
We consider a global set of aspects.

\subsection{Syntax} \label{secsb: trapactionsyntax}
The Syntax of AspectKE is given by Tables \ref{tab:syn_Aspects} and \ref{klaim:syn_NetProcess} (the KLAIM syntax).

\begin{table}[t] \changefont
$$
\begin{ARRAY}{lrcl}
S \in {\bf System} &
S & ::= & \Let\
\overrightarrow{asp}\ 
\Letin\ N
\\[1ex]
asp  \in  {\bf Asp} &
asp & ::= &  \adefn{A}{cut} body 
\\
body \in {\bf Advice} &
body & ::= & \Case \ (cond) \ sbody \ ; \ body 
\ \mid\ sbody \\
& sbody & ::= & \Break\ \mid\ \Proceed
\\[1ex]
cut  \in  {\bf Cut} &
cut & ::= & \Lc :: ca
\\[1ex]
ca  \in  {\bf CAct} &
ca & ::= & \Out{\veck{\Lnc}}{\ell} \mid \In{\veck{\Lntc}}{\ell} \mid
   \Read{\veck{\Lntc}}{\ell}  \mid \Eval{X}{\ell}  \mid {\bf newloc}({\un})
\\[1ex]
cond \in {\bf BExp} & 
cond & ::= &
\ell_1 = \ell_2 \ 
\mid \  cond_1  \land  cond_2  \ \mid \  cond_1  \lor  cond_2 \ \mid \ \neg \ cond \ 
\\ 
&&  \mid & \Test{\veck{\Lnc}}{\Lb}\ 
 \mid \exists u \in set: cond \mid \forall u \in set: cond
\\[1ex]
set  \in  {\bf Set} &
set & ::= &            
\{ \ell \} \mid set \cap set \mid set \cup set        
\\[1ex]
\Lnc,\Lntc \ \in \ {\bf Loc} &
\Lnc & ::= & \ell \  \mid\ \un \hfill 
  \Lntc\  ::= \  \Lnt \mid\ \un   
\end{ARRAY}
$$
\caption{AspectKE Syntax - Aspects for Trapping Actions} \label{tab:syn_Aspects}
\end{table}

Table \ref{tab:syn_Aspects} introduces a system $S$ (in {\bf System}) that consists of a net $N$ and a sequence of global aspect declarations $\veck{asp}$.  An aspect  declaration (in {\bf Asp}) takes the  form $\adefn{A}{cut} body $: $A$ is the aspect name, 
and $body$ (in {\bf Advice}) is the advice to the trapped  action. Each action (the {\bf Act} in Table \ref{klaim:syn_NetProcess}) is a potential join point that can be intercepted by AspectKE's pointcut (in {\bf Cut}).

Moreover, $\un$ is introduced as a don't-care parameter in the cut version of actions, and in the \textbf{test} primitive of conditional expressions ({\bf BExp}). It can match any type of location used in the program. Note in the cut, the occurrence of $!u$ and  $u$ have different meaning from those of KLAIM; 
a plain variable in a pointcut can only match an actual location and banged (!) variables in the pointcut can only match against binding occurrences of variables, while the don't-care ($\un$) can match both in the join point.

Each aspect gives a unique run-time suggestion (either {\bf break} or  {\bf proceed}) which may depend on the evaluation of a conditional expression.  The suggestion {\bf break} suppresses the trapped action whilst {\bf proceed}  allows the trapped action to be executed. In case of multiple aspects  that trap an action, {\bf break} takes precedence over {\bf proceed}.  The primitive $\Test{\veck{\Lnc}}{\Lb}$ evaluates to $\TT$ if a tuple exists in the tuple space of $\ell$ which matches $\veck{\Lnc}$.  Besides basic boolean expressions, condition $cond$ also includes \emph{bounded} existential quantification and universal quantification -- this allows simple queries to the databases occurring in the nets.

In contrast to other aspect languages, the condition is part of the advice instead of being part of the
pointcut (being evaluated before 
intercepting a join point). Evaluating the condition after intercepting a join point allows a more natural
modelling of security policies.

\paragraph{Well-formedness of Cuts.}

In addition to the well-formedness conditions for
KLAIM, we require that the variables 
in a cut are pairwise distinct. 
We shall also impose that aspects are \emph{closed}: any free variable in the body is defined in the cut. Additionally, when \emph{!u} is used in a cut pattern,  \emph{u} should not be used in conditions except in the context of set expressions.
\begin{examples}\label{example_a1} 
To illustrate how aspects can be composed in AspectKE that work with the KLAIM program, the following simple aspect gives advice to the running example in section \ref{runningexample}.
$$  
\begin{array}{l}
$$\small  
\begin{array}{l}
{\sf A}_1^{out}[user::{\bf out}(\un ,data)@{\sf \ds}] \\
\qquad \triangleq  {\bf case}(data={\sf alicetext})\\
\qquad \qquad {\bf break};\\
\quad \qquad{\bf proceed} 
\end{array}
$$ 
\end{array}
$$
The aspect 
traps an {\bf out} action of processes running at location {\sf \ds} that attempt to send a tuple with two fields. If the actual value of the second field is equal to {\sf alicetext}, the aspect will break the execution of the action and its continuation process. Otherwise, the action continues. 	
\end{examples}


\subsection{Semantics}\label{sec:aspectke_semantics}

The base semantics is that of KLAIM (Section \ref{sec:klaim}) but now,
before executing an action (all actions in a KLAIM program are potential join points), we check to see if any aspect applies to the action and combine the advice of all applicable aspects.  Each advice is either
that the action be allowed to proceed or not.  We resolve possible conflicts
by ensuring that any aspect that disallows an
action has priority.  Aspects are applied in definition order
but, because aspects can only allow or disallow the join point to proceed, the order is
actually immaterial.

\begin{examples}\label{example_a2} 
Suppose we have a system that contains the same net as in running example of Section \ref{runningexample} and aspect ${\sf A}_1^{out}$ in Example \ref{example_a1}:

\textbf{let} $\small
\begin{array}{l}
{\sf A}_1^{out}[user::{\bf out}(\un ,data)@{\sf \ds}] \\
\qquad \triangleq  {\bf case}(data={\sf alicetext})\\
\qquad \qquad {\bf break};\\
\quad \qquad{\bf proceed} 
\end{array} 
$ \textbf{in}

$\small
\begin{array}{ll}	
& {\sf \EDB} :: \langle {\sf Alice}, {\sf MedicalRecord}, {\sf \dn}, {\sf \PA}, {\sf alicetext} \rangle  \\
\netpar
& {\sf \EDB} :: \langle {\sf Bob}, {\sf PrivateNote}, {\sf \dsn}, {\sf \RE}, {\sf bobtext} \rangle  \\
\netpar
& {\sf \ds} :: \ {\bf read}({\sf Alice}, {\sf MedicalRecord}, {\sf \dn}, {\sf \PA},!content)@{\sf \EDB}.\\
& \qquad \qquad  \ {\bf out}({\sf Alice},content)@{\sf \ds}.\\ 
& \qquad \qquad  \ {\bf out}({\sf Alice}, {\sf MedicalRecord}, {\sf \ds}, {\sf \RE},{\sf newtext})@{\sf \EDB}\\
\end{array}
$

and some steps of execution (omitting the aspect definition):

$$\small
\begin{array}{ll}
& {\sf \EDB} :: \langle {\sf Alice}, {\sf MedicalRecord}, {\sf \dn}, {\sf \PA}, {\sf alicetext} \rangle  \\
\netpar
& {\sf \EDB} :: \langle {\sf Bob}, {\sf PrivateNote}, {\sf \dsn}, {\sf \RE}, {\sf bobtext} \rangle  \\
\netpar
& {\sf \ds} :: \ {\bf read}({\sf Alice}, {\sf MedicalRecord}, {\sf \dn}, {\sf \PA},!content)@{\sf \EDB}.\\
& \qquad \qquad  \ {\bf out}({\sf Alice},content)@{\sf \ds}.\\ 
& \qquad \qquad  \ {\bf out}({\sf Alice}, {\sf MedicalRecord}, {\sf \ds}, {\sf \RE},{\sf newtext})@{\sf \EDB}\\

\rightarrow\\
 & {\sf \EDB} :: \langle {\sf Alice}, {\sf MedicalRecord}, {\sf \dn}, {\sf \PA}, {\sf alicetext} \rangle  \\
 \netpar
& {\sf \EDB} :: \langle {\sf Bob}, {\sf PrivateNote}, {\sf \dsn}, {\sf \RE}, {\sf bobtext} \rangle  \\
 \netpar
&  {\sf \ds} :: \ {\bf out}({\sf Alice},{\sf alicetext})@{\sf \ds}.\\ 
& \qquad \qquad  \ {\bf out}({\sf Alice}, {\sf MedicalRecord}, {\sf \ds}, {\sf \RE},{\sf newtext})@{\sf \EDB}\\
\rightarrow\\
 & {\sf \EDB} :: \langle {\sf Alice}, {\sf MedicalRecord}, {\sf \dn}, {\sf \PA}, {\sf alicetext} \rangle  \\
\netpar
& {\sf \EDB} :: \langle {\sf Bob}, {\sf PrivateNote}, {\sf \dsn}, {\sf \RE}, {\sf bobtext} \rangle  \\
 \netpar
& {\sf \ds} :: \textbf{0}
\end{array}
$$

Aspect ${\sf A}_1^{out}$ does not trap the {\bf read} action, thus the {\bf read} action  executes and binds \emph{content} with {\sf alicetext}. But ${\sf A}_1^{out}$ traps the first {\bf out} action, and the result substitution is 
$$\small
\begin{array}{l}
[{\sf \ds}/user, {\sf alicetext}/data]
\end{array}
$$
and the case condition $data={\sf alicetext}$ evaluates to {\bf tt}, thus the aspect breaks the execution of this action and its continuation process.  		
\end{examples}


\section{Worked Examples: Advice for Access Control Models} \label{sec:access}

To evaluate the expressiveness of the language and show its language features,  we now show how AspectKE can be used to enforce access control policies by utilizing three well-known access control models, namely discretionary access control (Section \ref{subsec:dac}),  mandatory access control (Section \ref{subsec:mac}) and role-based access control (Section \ref{subsec:rac}), and how AspectKE can introduce new aspects for retrofitting new policies to existing systems (Section \ref{sec:multi}). 

Since patient confidentiality is an important issue in the health care industry it is imperative that EHRs are protected \cite{win2005rse}. To help achieve this goal, governments define many types of security policies, encapsulated in various acts and guides (e.g. \cite{NHS03,NHSCode}). Throughout the paper, we will enforce several security policies for the EHR system that was introduced in Section \ref{runningexample} and this shows different features of the language.  

The first is a primary use of data policy inspired by \cite{evered2004csa} which regulates the basic access control concerning the
\emph{read} and \emph{write} rights owned by doctors and nurses: 
\begin{qquote}
{\it Doctors can read all patients' medical records and private notes, while nurses can read all patients' medical records but cannot read 
any private notes. Medical records and private notes can only be created by doctors.}
\end{qquote}

For simplicity, here we restrict ourselves to only focussing on \textbf{read, in} and \textbf{out} actions, while \textbf{eval} and \textbf{newloc} actions will be  discussed further when enforcing other security policies. 

\subsection{Discretionary Access Control}\label{subsec:dac}
We will show how to enforce the above policy with discretionary access control (DAC), which is a type of access control as a means of restricting access to objects based on the identity of subjects and/or the groups to which they belong\cite{qiu1985trusted}.  We do so by using an \emph{access control matrix} containing triples $(s,o,c)$ identifying which subjects $s$ can perform which operations $c$ on which objects $o$. If we use the KLAIM programming model, we should equip the semantics of KLAIM with a reference monitor that consults the access control matrix when an action is executed to check if the action is permitted. In AspectKE we can directly use aspects to elegantly \emph{inline} the reference monitor to enforce this discretionary access control policy.

\begin{examples}\label{example3}
The access control matrix is stored in location {\sf DAC}, which  contains tuples:  $\langle user, recordtype, capability \rangle$. For example, if {\sf \ds} is a doctor and {\sf \no} is a nurse, then {\sf DAC} might contain the following tuples:
$$\small  
\begin{array}{l}
\langle {\sf \ds},{\sf MedicalRecord},{\sf read}\rangle\\
\langle {\sf \ds},{\sf PrivateNote},{\sf read}\rangle\\
\langle {\sf \ds},{\sf MedicalRecord},{\sf out}\rangle\\
\langle {\sf \ds},{\sf PrivateNote},{\sf out}\rangle\\
\langle {\sf \no},{\sf MedicalRecord},{\sf read}\rangle
\end{array}
$$ 
We also assume that the location {\sf DAC} can only be modified by privileged users, thus doctors and nurses cannot perform any {\bf in} and {\bf out} action on it. This can be enforced by other aspects but we omit them here.

The following aspect declarations will impose the desired requirements.
$$
\begin{array}{l}
$$\small  
\begin{array}{l}
{\sf A}_{\sf p1_{A1}}^{read}[user::{\bf read}(\un ,recordtype,\un,\un,\un)@{\sf \EDB}] \\
\qquad \triangleq  {\bf case}({\bf test}(user, recordtype, {\sf read})@{\sf DAC})\\
\qquad \qquad {\bf proceed};\\
\quad \qquad{\bf break} 
\end{array}
$$ 
\\

%

$$
\begin{array}{l}
{\sf A}_{\sf p1_{A2}}^{in}[user::{\bf in}(\un ,recordtype,\un,\un,\un)@{\sf \EDB}] \\
\qquad \triangleq  {\bf case}({\bf test}(user, recordtype, {\sf in})@{\sf DAC})\\
\qquad \qquad {\bf proceed};\\
\quad \qquad{\bf break} 
\end{array}
$$ \\

$$\small
\begin{array}{l}
{\sf A}_{\sf p1_{A3}}^{out}[user::{\bf out}(\un,recordtype,\un,\un,\un)@{\sf \EDB}] \\
\qquad \triangleq  {\bf case}({\bf test}(user, recordtype, {\sf out})@{\sf DAC})\\
\qquad \qquad {\bf proceed};\\
\quad \qquad{\bf break} 
\end{array}
$$
\end{array}
$$
Aspects ${\sf A}_{\sf p1_{A1}}^{read}$, ${\sf A}_{\sf p1_{A2}}^{in}$, and ${\sf A}_{\sf p1_{A3}}^{out}$  enforce the above policy by using \textsf{DAC}, where the access rights for each user are actually described. Note that the second field of the tuple operated by these cut actions is \emph{recordtype}, which trap an action that clearly specifies a concrete record type. 

Consider the following KLAIM program that is a variant of the running example in Section \ref{runningexample} (in that the user is  nurse {\sf \no} instead of doctor {\sf \ds}) and is equipped with the above four aspects:
$$\small
\begin{array}{ll}
{\sf \no} :: & {\bf read}({\sf Alice,\MR, \dn,\PA},!content)@{\sf \EDB}.\\ 
& {\bf out}({\sf Alice},content)@{\sf \no}.\\
&{\bf out}({\sf Alice,\MR,\no,\RE,newtext})@{\sf \EDB}
\end{array}
$$ 
The first {\bf read} action will be trapped by aspect ${\sf A}_{\sf p1_{A1}}^{read}$, and the resulting substitution is
$$\small
\begin{array}{l}
[{\sf \no}/user, {\sf \MR}/recordtype]
\end{array}
$$
and the condition {\bf test}({\sf \no}, {\sf MedicalRecord}, {\sf read})@{\sf DAC} is evaluated.  Since {\sf \no} has the appropriate right according to {\sf DAC} we proceed and perform this {\bf read} action thereby giving rise to the binding of $content$ to {\sf alicetext}.

The second action will not be trapped by any of the aspects, so it will simply be performed and the tuple 
$\langle {\sf Alice}, {\sf alicetext} \rangle$ is output to location {\sf \no}.

The last action will be trapped by aspect ${\sf A}_{\sf p1_{A3}}^{out}$ and after the substitution we evaluate the condition {\bf test}({\sf \no}, {\sf MedicalRecord}, {\sf out})@{\sf DAC} which is evaluated to {\FF} and thus we {\bf break} the execution.

However, the KLAIM program can also execute \textbf{read} or \textbf{in} actions without specifying the record type, e.g., using !\emph{recordtype} instead of \emph{recordtype}, users can thus get a record as follows:
$$\small {\sf \no} :: {\bf read}({\sf Alice,{!\mathit{recordtype}}, \dn,\PA},!content)@{\sf \EDB}
$$
where a successful input action can retrieve any type of EHR record.

None of the above aspects can trap these input actions, thus we have to enforce additional aspects so that the above input actions will not bypass our aspects and consequently break the policy. The simple aspects forbid any attempts to \textbf{read} or \textbf{in} (read and then remove) EHR records without specifying the record type: 
$$  
\begin{array}{l}
$$\small
\begin{array}{l}
{\sf A}_{\sf p1_{A4}}^{read}[user::{\bf read}(\un ,!recordtype,\un,\un,\un)@{\sf \EDB}] \\
\qquad \triangleq {\bf break} 
\end{array}
$$ \\

$$\small
\begin{array}{l}
{\sf A}_{\sf p1_{A5}}^{in}[user::{\bf in}(\un,!recordtype,\un,\un,\un)@{\sf \EDB}] \\
\qquad \triangleq {\bf break} 
\end{array}
$$
\end{array}
$$
One may wonder why not build the above two aspects on top of aspects ${\sf A}_{\sf p1_{A1}}^{read}$ and ${\sf A}_{\sf p1_{A2}}^{in}$ by directly replacing \emph{recordtype} with !\emph{recordtype} in their pointcut, respectively. The reason is that these aspects will not be well-formed: when trapping actions, \emph{recordtype} binds with a variable, which cannot be used in a test condition such as  ${\bf test}(user, recordtype, {\sf read})@{\sf DAC}$. 
\end{examples}

\subsection{Mandatory Access Control}\label{subsec:mac}
In this subsection we will show how to enforce the above policies by using mandatory access control (MAC), which is a means of restricting access to objects based on the sensitivity (as represented by a label) of the information contained in the objects and the formal authorization (i.e., clearance) of subjects\cite{qiu1985trusted}. Before enforcing the above policy, we first impose a comparable classical MAC policy - the Bell-LaPadula security policy \cite{Gollmann199902} based on a mandatory access control model. Later we enforce the above policy as a variant of the Bell-LaPadula policy. In the presentation, \emph{security levels} are assigned to subjects (as clearances) and objects (as labels). 

\begin{examples}\label{example4}
In this scenario, we just need two security levels, and may assign security levels to subjects as follows: doctors have level {\sf high} and nurses have level {\sf low}; similarly we may assign objects as follows: private notes have level {\sf high} and medical records have level {\sf low}. 

To model this policy we need to introduce a location {\sf MAC} that stores tuples of the form: 
$\langle user, securitylevel\rangle$ and $\langle recordtype, securitylevel\rangle$. Continuing Example \ref{example3}, we create the location {\sf MAC} with the tuples:
$$\small  
\begin{array}{l}
\langle {\sf \ds},{\sf high}\rangle\\
\langle {\sf \no},{\sf low}\rangle\\
\langle {\sf PrivateNote},{\sf high}\rangle\\
\langle {\sf MedicalRecord},{\sf low}\rangle\\
\end{array}
$$ 
As before we also assume that the location {\sf MAC} can only be modified by privileged users.

Firstly, we enforce the Bell-LaPadula security policy \cite{Gollmann199902} to illustrate that AspectKE can enforce a well-known mandatory access control policy.  Then we will enforce our example policy, with small modifications based on the aspects that enforce Bell-LaPadula policy. 

If we enforce the Bell-LaPadula security policy, the first part of the policy states that a subject is allowed to read or input data from any object provided that the subject's security level dominates that of the object. 
In our case, this guarantees \emph{no read up}:  that is,  {\sf low} subjects (nurses) cannot read high objects (private notes) but can only read low objects (medical records); however, {\sf high} subjects (doctors) can access both kinds of records.

The \emph{no read up} part of the policy can be enforced by aspects as follows:
$$
\begin{array}{l}
$$\small
\begin{array}{l}
{\sf A}_{\sf p1_{B1}}^{read}[user::{\bf read}(\un,recordtype,\un,\un,\un)@{\sf \EDB}] \\
\qquad \triangleq  {\bf case}(\lnot ({\bf test}(user, {\sf low})@{\sf MAC} \land {\bf test}(recordtype, {\sf high})@{\sf MAC}  ))\\
\qquad \qquad {\bf proceed};\\
\quad \qquad{\bf break} 
\end{array}
$$ 
\end{array}
$$
The second part of the policy (a simplified form of Bell-LaPadula star property \cite{Gollmann199902}) states that a subject can write to any object provided that the security level of the object dominates that of the subject (\emph{no write down}). In our case {\sf high} subjects (doctors) cannot write low objects (medical records) but {\sf low} subjects (nurses) can write to both kinds of records.

The \emph{no write down} of the policy can be enforced by the aspect below:
$$\small
\begin{array}{l}
{\sf A}_{\sf p1_{B2}}^{out}[user::{\bf out}(\un,recordtype,\un,\un,\un)@{\sf \EDB}] \\
\qquad \triangleq  {\bf case}(\lnot ({\bf test}(user, {\sf high})@{\sf MAC} \land {\bf test}(recordtype, {\sf low})@{\sf MAC}  ))\\
\qquad \qquad {\bf proceed};\\
\quad \qquad{\bf break} 
\end{array}
$$
Additionally, we have an aspect for the \textbf{read} action to prevent users from reading records without specifying the record type, and an aspect for
the \textbf{in} action to prevent users from reading and deleting records:
$$
\begin{array}{l}
$$\small
\begin{array}{l}
{\sf A}_{\sf p1_{B3}}^{read}[user::{\bf read}(\un,!recordtype,\un,\un,\un)@{\sf \EDB}] \\
\qquad \triangleq {\bf break} 
\end{array}
$$\\

$$\small
\begin{array}{l}
{\sf A}_{\sf p1_{B4}}^{in}[user::{\bf in}(\un,\un,\un,\un,\un)@{\sf \EDB}] \\
\qquad \triangleq {\bf break} 
\end{array}
$$
\end{array}
$$
These aspects correctly enforce our policy about reading patient records.
However, the no write down policy is not quite right for our example, instead we depart from the Bell-LaPadula policy and define:
$$\small
\begin{array}{l}
{\sf A}_{\sf p1_{B2'}}^{out}[user::{\bf out}(\un,recordtype,\un,\un, \un)@{\sf \EDB}] \\
\qquad \triangleq  {\bf case}({\bf test}(user, {\sf high})@{\sf MAC} )\\
\qquad \qquad {\bf proceed};\\
\quad \qquad{\bf break} 
\end{array}
$$
This aspect allows doctors to write any kind of record.

The aspect ${\sf A}_{\sf p1_{B2'}}^{out}$ together with ${\sf A}_{\sf p1_{B1}}^{read},{\sf A}_{\sf p1_{B3}}^{read}, {\sf A}_{\sf p1_{B4}}^{in}$ reflect a mandatory access control model which satisfies our policy. In this case we only allow {\sf high} users (doctors) to write patient records. Hence nurse {\sf \no} in Example \ref{example3} cannot execute the third action as it will be blocked by ${\sf A}_{\sf p1_{B2'}}^{out}$, which would be allowed with ${\sf A}_{\sf p1_{B2}}^{out}$ from the Bell-LaPadula security policy.
\end{examples}

\subsection{Role-Based Access Control}\label{subsec:rac}
Role-based access control (RBAC) \cite{sandhu1996rba} is another access control mechanism which allows  the central administration of security policies and is often more flexible and elegant for modelling  security policies. The simplest model in the RBAC family is $RBAC_0$, where there are three sets of entities called \emph{user}, \emph{role}, and \emph{permission}. A user can be assigned multiple roles (role assignment) and a role can have multiple permissions (permission assignment) to corresponding operations. In addition, the user can initiate a \emph{session} during which the user activates some subset of roles that he or she has been assigned. A user can execute an operation only if the user's active roles have the permission to perform that operation.

\begin{examples}\label{example5}
To implement the security policy for patient records, we use a model that does not differentiate a user's assigned role and active role (we assume that the assigned roles of all users are activated by default), so we only need location {\sf \ER} with tuples $\langle user, role \rangle$:
$$\small  
\begin{array}{l}
\langle {\sf \ds},{\sf Doctor}\rangle\\
\langle {\sf \no},{\sf Nurse}\rangle
\end{array}
$$
For permission assignment we also need a location to describe each role's permission. This can be done by storing tuples  
$\langle role, object, capability \rangle$ at ${\sf \EP}$: 
$$\small  
\begin{array}{l}
\langle {\sf Doctor, MedicalRecord},{\sf read}\rangle\\
\langle {\sf Doctor, PrivateNote},{\sf read}\rangle\\
\langle {\sf Doctor, MedicalRecord},{\sf out}\rangle\\
\langle {\sf Doctor, PrivateNote},{\sf out}\rangle\\
\langle {\sf Nurse, MedicalRecord},{\sf read}\rangle
\end{array}
$$
Once more we assume that the locations {\sf RDB} and {\sf PDB} can only be modified by privileged users.

The following aspects then implement the required policy:
$$
\begin{array}{l}
$$\small
\begin{array}{l}
{\sf A}_{\sf p1_{C1}}^{read}[user::{\bf read}(\un,recordtype,\un,\un,\un)@{\sf \EDB}] \\
\qquad \triangleq  {\bf case}(
\exists role \in \{{\sf Doctor, Nurse} \}:\\
\qquad \qquad \qquad ({\bf test}(user, role)@{\sf \ER} \land {\bf test} (role, recordtype, {\sf read} )@{\sf \EP}))\\
\qquad  \qquad {\bf proceed};\\
\qquad \quad {\bf break} 
\end{array}
$$	
\\

%

$$\small
\begin{array}{l}
{\sf A}_{\sf p1_{C2}}^{in}[user::{\bf in}(\un,recordtype,\un,\un,\un)@{\sf \EDB}] \\
\qquad \triangleq  {\bf case}(
\exists role \in \{{\sf Doctor, Nurse} \}:\\
\qquad \qquad \qquad ({\bf test}(user, role)@{\sf \ER} \land {\bf test} (role, recordtype, {\sf in} )@{\sf \EP}))\\
\qquad  \qquad {\bf proceed};\\
\qquad \quad {\bf break} 
\end{array}
$$	
\\

$$\small
\begin{array}{l}
{\sf A}_{\sf p1_{C3}}^{out}[user::{\bf out}(\un,recordtype,\un,\un,\un)@{\sf \EDB}] \\
\qquad \triangleq  {\bf case}(
\exists role \in \{{\sf Doctor, Nurse} \}:\\
\qquad \qquad \qquad ({\bf test}(user, role)@{\sf \ER} \land {\bf test} (role, recordtype, {\sf out} )@{\sf \EP}))\\
\qquad  \qquad {\bf proceed};\\
\qquad \quad {\bf break} 
\end{array}
$$

\end{array}
$$

These three aspects are useful for interrupting the execution when a user attempts to operate on EHR records with a concrete record type, which essentially relies on the tuples from \textsf{RDB} and \textsf{PDB}. They also show the benefit of admitting quantifiers into the conditional expressions.

Similar to the previous subsections, we have to enforce additional aspects for capturing user attempts to access EHR records without specifying the record type. 
$$
\begin{array}{l}
$$\small
\begin{array}{l}
{\sf A}_{\sf p1_{C4}}^{read}[user::{\bf read}(\un,!recordtype,\un,\un,\un)@{\sf \EDB}] \\
\qquad \triangleq  {\bf break} 
\end{array}
$$
\\

$$\small
\begin{array}{l}
{\sf A}_{\sf p1_{C5}}^{in}[user::{\bf in}(\un,!recordtype,\un,\un,\un)@{\sf \EDB}] \\
\qquad \triangleq {\bf break} 
\end{array}
$$
\end{array}
$$
In the following sections we will use the role-based access control mechanism since it is best suited for enforcing security policies in a large organization. 
\end{examples} 
{

\subsection{Advice for Retrofitting Policies}\label{sec:multi}

Now we will show how aspects in AspectKE can retrofit new security policies into an exiting system that is being developed/updated or has already been deployed. Concretely, when a new functionality has been introduced to the existing system we will show how we enforce new policies to cater for the new requirements (Section \ref{subsubsec:refining1}); when a policy has been proposed to refine existing policies, we will show how we enforce policies based on the existing functionality of the system (Section \ref{subsubsec:refining2}); on the other hand,  sometimes additional functionality has to be introduced to the system to implement aspects which refine existing policies (Section \ref{subsubsec:refining3}). 

Indeed, the possibility to refine, renew and retrofit security policies into an existing/evolving system is very important for IT systems. Taking the EHR system as an example, as the public debate about security standards (especially for privacy) evolves, governments have to modify the corresponding acts and guides.  As a consequence, the security policy for an EHR system will undergo frequent change \cite{becker2004cft}. Moreover, the  IT system itself will always be enhanced by new functionalities, which means that new policies need to be enforced. The National IT Strategy for the Danish Health Care Service states that ``it is also important to acknowledge the fact that IT is not just implemented once and for all''\cite{danishhc}. 

\subsubsection{Enforcing Security Policy for New Functionality}\label{subsubsec:refining1}

AspectKE can be used for enforcing new security policies when a new functionality has been introduced at any phase of the system development. The running example in Section \ref{runningexample} introduces the functionality of the EHR system as regards reading and writing from and to the EHR database ({\sf \EDB}). Now we introduce another (new) function to the EHR system that enables a manager to add a patient, or delete information from the system. 

In our programming model, each patient is represented by a location. A manager can add a new patient to the system as follows:
$$\small
\begin{array}{ll}
{\sf \md} :: & {\bf newloc}(!patient).{\bf out}(patient, {\sf Patient})@{\sf \ER} 
\end{array}
$$ 
First a new location for the  patient is created, then it is registered in the role database {\sf \ER} by the {\bf out} action. To delete a user one can simply perform an {\bf in} action to the location {\sf RDB}.

We shall now show how to enforce the following security policy regarding the manager role at the hospital\cite{evered2004csa}, for this new added functionality. 

\begin{qquote}
{\it In the hospital, only managers are allowed to add a user to, or delete from the system.}
\end{qquote}
 
\begin{examples} \label{example8}
The following aspects will enforce the above security policy.
$$\small
\begin{array}{l}	
\begin{array}{l}
{\sf A}_{\sf p2}^{newloc}[user::{\bf newloc}(\un)] \\
\qquad \triangleq  {\bf case}(
 {\bf test}(user, {\sf Manager})@{\sf \ER} \land {\bf test} ({\sf Manager}, {\sf Location}, {\sf newloc} )@{\sf \EP})\\
\qquad  \qquad {\bf proceed};\\
\qquad \quad {\bf break}
\end{array}
\\
\begin{array}{l}
{\sf A}_{\sf p2}^{out}[user::{\bf out}(\un,\un)@{\sf \ER}] \\
\qquad \triangleq  {\bf case}(
{\bf test}(user, {\sf Manager})@{\sf \ER} \land {\bf test} ({\sf Manager}, {\sf \ER}, {\sf out} )@{\sf \EP})\\
\qquad  \qquad {\bf proceed};\\
\qquad \quad {\bf break} 
\end{array}
\\
\begin{array}{l}
{\sf A}_{\sf p2}^{in}[user::{\bf in}(\un,\un)@{\sf \ER}] \\
\qquad \triangleq  {\bf case}(
{\bf test}(user, {\sf Manager})@{\sf \ER} \land {\bf test} ({\sf Manager}, {\sf \ER}, {\sf in} )@{\sf \EP})\\
\qquad  \qquad {\bf proceed};\\
\qquad \quad {\bf break} 
\end{array}
\end{array}
$$
These aspects are composed in a way that is similar to the Example \ref{example5}. Before using them we need to put the tuple {$\langle{\sf \md},{\sf Manager}\rangle$}  into {\sf \ER} and the tuples {$\langle{\sf Manager}, {\sf Location}, {\sf newloc}\rangle$},  {$\langle{\sf Manager}, {\sf \ER}, {\sf out}\rangle$}, and {$\langle{\sf Manager}, {\sf \ER}, {\sf in}\rangle$} into {\sf \EP}. Finally, once again we assume that {\sf PDB} can only be modified by privileged users. 
\end{examples}

Example \ref{example8} shows that AspectKE can enforce a policy for new functionality (code), no matter when this functionality is developed, potentially in the entire life cycle of the EHR system. This is because the underlying aspect-oriented mechanism allows new aspects to intercept all join points (including a join point of new code) and give appropriate advice.  Example \ref{example8} also shows how to give advice on an action ({\bf newloc}) that creates new node in a net.
Note that Examples \ref{example5} and \ref{example8} use the role-based access control model which shows that the tuples introduced above at {\sf \EP} are good at expressing permissions that only directly rely on roles and objects. However, some permission assignments are more complex and therefore we shall also use logical formulae to express permission assignments in the following sections.

\subsubsection{Refining Security Policy with Existing Functionality}\label{subsubsec:refining2}

AspectKE can be used to to refine an existing security policy at any phase of the system development. The following is a policy which can be considered as a  refinement of the previous policy (enforced by Example \ref{example3}-\ref{example5}) to protect patients' privacy. This policy is also inspired by \cite{evered2004csa}:
\begin{qquote}
{\it Private notes can only be viewed on the basis of the 
doctor-patient confidentiality -- 
doctors cannot view private notes that were not created by themselves; 
nurses can only view recent medical records which were created after 
the patient has 
been admitted.}
\end{qquote}

Traditional programming paradigms normally necessitate modifying existing code  to enforce this extra policy, while the aspect approach simply requires additional  aspects.

\begin{examples}\label{example6}
	The first part of the  policy can be expressed by the aspects shown below. These two aspects will prevent a doctor from reading a private note written by another doctor or reading a private note without clearly specifying the author of the note. 
	$$
	\begin{array}{l}
	$$\small
	\begin{array}{l}
	{\sf A}_{\sf p3_1}^{read}[user::{\bf read}(\un,{\sf \PN},author,\un,\un)@{\sf \EDB}] \\
	\qquad \triangleq  {\bf case}( {\bf test}(user, {\sf Doctor})@{\sf \ER} \land  (user=author))\\
	\qquad \qquad {\bf proceed};\\ 
	\qquad \quad  {\bf case}(\lnot {\bf test}(user, {\sf Doctor})@{\sf \ER} ) \\
	\qquad \qquad {\bf proceed};\\ 
	\qquad \quad {\Break} 
	\end{array}
	$$

	\\

	$$\small
	\begin{array}{l}
	{\sf A}_{\sf p3_2}^{read}[user::{\bf read}(\un,{\sf \PN},!author,\un,\un)@{\sf \EDB}] \\
	\qquad \triangleq  {\bf case}(\lnot {\bf test}(user, {\sf Doctor})@{\sf \ER} ) \\
	\qquad \qquad {\bf proceed};\\ 
	\qquad \quad {\Break}
	\end{array}
	$$

	\end{array}
	$$	
	Note that the second case of ${\sf A}_{\sf p3_1}^{read}$ and 
${\sf A}_{\sf p3_2}^{read}$ allow any users registered in {\sf RDB} except doctors to read a private note, which includes users taking roles as nurses. Allowing nurses to read a private note is not problematic, as these two aspects only reflect the intention of this policy. Preventing nurses to read a private note is enforced in the previous policy. 
	
	These two aspects are supposed to work together with those aspects defined in the Examples \ref{example3} to \ref{example5}. For example if a nurse tries to read a private note, these aspects will {\bf proceed} the execution but aspects in Examples \ref{example3}-\ref{example5} will suggest {\bf break}.   
	 Another example is if a doctor tries to read a private note written by other doctors, aspects in Examples \ref{example3}-\ref{example5} will allow the action to {\bf proceed} whilst these aspects will {\bf break} the execution. Since {\bf break} takes precedence over {\bf proceed}, the final decision suggested from this advice will be to block the execution in both cases.

The second part of the policy says that nurses can only read recent medical records which can be expressed by the following aspects where, once again, the aspects should be considered in conjunction with those of Examples \ref{example3} to \ref{example5}. These two aspects will prevent a nurse from reading a past medical record or reading a medical record without specifying the created time.
$$
\begin{array}{l}
$$\small
\begin{array}{l}
{\sf A}_{\sf p3_3}^{read}[user::{\bf read}(\un,{\sf \MR},\un,createdtime,\un)@{\sf \EDB}] \\
\qquad \triangleq {\bf case}( {\bf test} (user, {\sf Nurse})@{\sf \ER} \land  createdtime={\sf \RE})\\
\qquad \qquad {\bf proceed};\\ 
\qquad \quad {\bf case}(\lnot{\bf test}(user, {\sf Nurse})@{\sf \ER})\\
\qquad \qquad {\bf proceed};\\ 
\qquad \quad {\Break}\end{array}
$$
\\
$$\small
\begin{array}{l}
{\sf A}_{\sf p3_4}^{read}[user::{\bf read}(\un,{\sf \MR},\un,!createdtime,\un)@{\sf \EDB}] \\
\qquad \triangleq {\bf case}(\lnot{\bf test}(user, {\sf Nurse})@{\sf \ER})\\
\qquad \qquad {\bf proceed};\\ 
\qquad \quad {\Break}\end{array}
$$
\end{array}
$$
\end{examples}
Note that no new functionality is required to enforce this policy, as these aspects only rely on the existing program (i.e., node {\sf RDB}).

\subsubsection{Refining Security Policy with New Functionality}\label{subsubsec:refining3}

Sometimes refining an existing security policy is necessary. We have to introduce additional functionality to support the implementation of the aspects for policy refinement.
Now we consider the following security policy that restricts access to the database to certain locations in the hospital \cite{beznosov1998rac}:
\begin{qquote}
{\it A nurse can only read medical records of the patients who are in 
the wards located on the nurse's working floor. Furthermore, the nurse 
can only access 
medical records through the computers that are located on that specific 
floor. But in the emergency room, a nurse does not have this restriction.}
\end{qquote} 

\begin{examples}\label{example7}
To express this policy as aspects we shall assume that the current location database {\sf \ECL} records  every user's current location information (indicating that they are using computers at that location), and that the assigned location database {\sf \EAL} stores every user's assigned room (e.g. for nurses this is their working floors and rooms which they are responsible for; for patients this is the floors and wards that they are on). These two databases store tuples with the same fields $\langle user,floor,room \rangle$ and can only be modified by privileged users.

The set {\sf Floor} contains the actual floors of the hospital (e.g. {\sf f1},{\sf f2}). The set {\sf Room} contains the actual rooms of the hospital and includes two types of rooms: ordinary wards (e.g. {\sf w1},{\sf w2}) and the special room {\sf EmergencyRoom}. 

The appropriate advice can now be expressed as follows:
$$\begin{array}{l}
$$\small
\begin{array}{l}
{\sf A}_{\sf p4_1}^{read}[user::{\bf read}(patient,{\sf \MR},\un,\un,\un)@{\sf \EDB}] \\
\qquad \triangleq  {\bf case}({\bf test}(user, {\sf Nurse})@{\sf \ER} \land \\
\qquad \qquad \ \ \ \  \exists floor \in {\sf Floor:} \ {\bf test}(user, floor,{\sf EmergencyRoom})@{\sf \ECL})\\
\qquad \qquad {\bf proceed};\\
\qquad \quad  {\bf case}({\bf test}(user, {\sf Nurse})@{\sf \ER} \land  \\
\qquad \qquad \ \ \ \  \exists floor\in{\sf Floor}\ \exists room \in{\sf Room}:(
\begin{array}[t]{l} {\bf test}(user,floor, room)@{\sf \ECL}  \land \\
{\bf test}(user,floor,room)@{\sf \EAL} \land \\
{\bf test}(patient,floor,room)@{\sf \EAL}))
\end{array}\\
\qquad \qquad {\bf proceed};\\
\qquad \quad  {\bf case} (\lnot {\bf test}(user, {\sf Nurse})@{\sf \ER})\\
\qquad \qquad {\bf proceed};\\ 
\qquad \quad {\Break}
\end{array}
$$

\\

$$\small
\begin{array}{l}
{\sf A}_{\sf p4_2}^{read}[user::{\bf read}(!patient,{\sf \MR},\un,\un,\un)@{\sf \EDB}] \\
\qquad \triangleq  {\bf case}({\bf test}(user, {\sf Nurse})@{\sf \ER} \land \\
\qquad \qquad \ \ \ \  \exists floor \in {\sf Floor:} \ {\bf test}(user, floor,{\sf EmergencyRoom})@{\sf \ECL})\\
\qquad \qquad {\bf proceed};\\
\qquad \quad  {\bf case} (\lnot {\bf test}(user, {\sf Nurse})@{\sf \ER})\\
\qquad \qquad {\bf proceed};\\ 
\qquad \quad {\Break}
\end{array}
$$
\end{array}$$
In ${\sf A}_{\sf p4_1}^{read}$, the first case caters for the situation where a nurse is working in the emergency room; the second case allows the {\bf read} action when a nurse is trying to access the medical record for a patient who is on the same ward/floor as where the nurse is currently at and assigned to work, the third case allows a user who is not a nurse to perform the {\bf read} action. The aspect ${\sf A}_{\sf p4_2}^{read}$ is similar to the aspect ${\sf A}_{\sf p4_1}^{read}$ except that it does not contain the second case. This is because reading a medical record without specifying the name of the patient is not acceptable for a nurse who is not working in the emergency room.

As in Example \ref{example6}, these aspects will work together with all previously introduced security polices. Moreover, these aspects are both in the spirit of role-based access control, and they demonstrate that when composing aspects in AspectKE for larger and more complex security policies of an organization, role-based access control can be very efficacious, as has already been observed in the literature \cite{sandhu1996rba}.
\end{examples}
Note new functionality has to be introduced to enforce this policy such as the newly introduced nodes for databases (e.g., {\sf CLDB, ALDB}). This new functionality can be developed as part of the main logic of the EHR system, or can be merely developed and maintained to enforce security policies. We observe that although we might need to extend the functionality of the system's main logic to enforce certain security policies,  the policies themselves are still described in aspects (even though they directly or indirectly rely on new nodes/processes), which are still separated  out of the main logic.  






\section{AspectKE: Trapping Processes} \label{sec:process}

\subsection{Motivation}
Classical reference monitors are incapable of enforcing security policies based on the future behavior of programs, rather they rely only on information gathered by monitoring execution steps \cite{schneider2000esp}, and perform history-based dynamic checks. However, security polices concerned with information flow that cannot be implemented correctly without a security check of the overall behavior of the program exist. 

For example, in a software system, remote evaluation involves the transmission of programs from a client to a server for subsequent execution at the server. However, as the programs transmitted might perform unintended operations at the server side, a security check is usually needed.  A typical example of this is Java applets  which can be transferred to a remote system and executed by the Java Virtual Machine (JVM). Since the unknown applets are not always written by trusted users, the JVM has  certain mechanisms for ensuring that the applet will not be able to do malicious actions, e.g., the bytecode verifier \cite{leroy2001java} and sandbox mechanism \cite{oaks1998java}. 

As another example, in the EHR domain, rather than enforcing the primary use of data policies for direct patient care domain as shown in the previous sections, there is a trend to define and enforce secondary use of data policies. Here data is used outside of direct health care delivery that includes activities such as analysis, research,  public health etc,  even though it still lacks a robust infrastructure of policies and is surrounded with complex ethical, political, technical and  social issues \cite{safran2007tnf}.  Compared with primary use of data, whose focus is on regulating \emph{``someone has some rights to access some data''}, it focuses on defining \emph{"how the data can be used after it is released to someone"}. The enforcement of such policies requires security checks in the form of inspection of the flow of data.

In general, program analysis techniques concern computing reliable, approximate information about the dynamic behavior of programs \cite{NNH_PPA}, and the derivation of useful information by simulating execution of all possible paths of the executing program. For example, type systems can be used to enforce various kinds of information flow as well as access control security policies (like Jif \cite{myers2001jji} for Java). 

In this section, we extend the AspectKE language presented in Section \ref{sec:review} which enables AspectKE not only to trap the action, but also to trap a process that is to be executed in the future. The process can be a process that is to be sent to a remote site, or a process continuation of a trapped action. This enables us to perform simple forms of program analyses, called \emph{behavior analysis}, that syntactically inspect the trapped processes, i.e., actions in a new thread to be executed (at local/remote sites) or the remaining actions in the current thread.  In the following section, we will show how to use simple behavior analysis techniques to enforce various security policies that require checking of the future behavior of a program, the so-called predictive access control policies, and that detect and prevent execution of the potential malicious operations at the earliest stage.


\subsection{Extended Syntax and Semantics} \label{subsec:extendedsyntax}

\begin{table}[hpt] \changefont
$$
\begin{ARRAY}{lrcl}
cut  \in  {\bf Cut} &
cut & ::= &\cdots \mid   \Lc :: ca.X
\\[1ex]
cond \in {\bf BExp} & 
cond & ::= &
\cdots \mid c \in set \mid \ell \in set \mid
           set = \emptyset 
\\[1ex]
set  \in  {\bf Set} &
set & ::= & \cdots  \mid \{ c \} \mid  {\sf Act}(X) \mid  {\sf Loc}_{c}(X) \mid
            {\sf FV}(X) \mid {\sf FV}_{c}(X) \mid {\sf LC}(X) \mid {\sf LC}_c(X)
\mid   {\sf LVar}_*    
\end{ARRAY}
$$
\caption{AspectKE Syntax - Aspects for Trapping Processes} \label{tab:syn_Aspects_process}
\end{table}



Table \ref{tab:syn_Aspects_process} shows the extended syntax of AspectKE. 

The pointcut ({\bf Cut}) in Table \ref{tab:syn_Aspects} has been extended with $\Lc::ca.X$, which not only binds the action, but also binds the program continuation after the trapped action to a variable $X$.

Table \ref{tab:syn_Aspects_process}  extends {\bf BExp} and {\bf set} in Table \ref{tab:syn_Aspects}, which can be used for defining properties that require syntactic analysis of the processes to be executed  (usually the continuation of the trapped action bound by $X$). The set-forming behavior analysis functions ${\sf Act},{\sf Loc}_c,{\sf LC},{\sf LC}_c,{\sf FV},{\sf FV}_c$ will be explained in the following sections when needed, but we have collected their definitions in Table \ref{PASet_functions} where $fv$, $bv$, $lc$, $cap$ and $loc$ are
the obvious extraction functions for free variables, bound variables, location constants, capabilities and target locations, respectively. 

Note these functions expose different data-flow information of processes bound by process variables, and can be used to enforce \emph{predictive access control policies}, namely access control policies that depend on the future behavior of a program.   


%


\begin{table}[t] \changefont
$$
\begin{array}{lll}
\hline
\\[-1ex]
{\sf Act} (P_1|P_2) &=& {\sf Act}(P_1) \cup {\sf Act}(P_2)\\
{\sf Act} ( \Sigma_i a_i.P_i) &=& \bigcup_i (\{{cap}(a_i)\} \cup {\sf Act}(P_i)) \\
{\sf Act}(*P) &=& {\sf Act}(P)
\\[1ex]
\hline

\\[-1ex]
{\sf Loc}_c (P_1|P_2) &=& {\sf Loc}_c (P_1) \cup {\sf Loc}_c (P_2)\\
{\sf Loc}_c ( \Sigma_i a_i.P_i) &=& \bigcup_i (
\{ {loc}(a_i) \mid {cap}(a_i) = c\}  \cup {\sf Loc}_c (P_i)) \\
{\sf Loc}_c (*P) &=& {\sf Loc}_c (P)
\\[1ex]

\hline

\\[-1ex]
{\sf LC} (P_1|P_2) &=& {\sf LC}(P_1) \cup {\sf LC}(P_2)\\
{\sf LC} ( \Sigma_i a_i.P_i) &=& \bigcup_i ({lc}(a_i) \cup {\sf LC}(P_i)) \\
{\sf LC}(*P) &=& {\sf LC}(P)
\\[1ex]

\hline

\\[-1ex]
{\sf LC}_c (P_1|P_2) &=& {\sf LC}_c(P_1) \cup {\sf LC}_c(P_2)\\
{\sf LC}_c ( \Sigma_i a_i.P_i) &=& \bigcup_i (\{{lc}(a_i)\mid {cap}(a_i)=c \} \cup {\sf LC}(P_i)) \\
{\sf LC}_c(*P) &=& {\sf LC}_c(P)
\\[1ex]

\hline

\\[-1ex]
{\sf FV} (P_1|P_2) &=& {\sf FV} (P_1) \cup {\sf FV}(P_2)\\
{\sf FV} ( \Sigma_i a_i.P_i) &=& \bigcup_i ( {fv}(a_i) \cup ({\sf FV}(P_i) \setminus bv(a_i)) )\\
{\sf FV} (*P) &=& {\sf FV} (P)
\\[1ex]
\hline


\\[-1ex]
{\sf FV}_c (P_1|P_2) &=& {\sf FV}_c (P_1) \cup {\sf FV}_c (P_2)\\
{\sf FV}_c ( \Sigma_i a_i.P_i) &=& \bigcup_i (  \{ {fv}(a_i) \mid {cap}(a_i)=c \} \cup 
 ({\sf FV}_c (P_i) \setminus bv(a_i))) \\
{\sf FV}_c (*P) &=& {\sf FV}_c (P)
\\[1ex]
\hline
\end{array}
$$

\caption{Behavior Analysis Functions } \label{PASet_functions}
\end{table}


\begin{examples}\label{example_b1}  
To illustrate how aspects in the extended AspectKE trap a KLAIM program and extract its properties, the following aspect gives advice to the running example in section \ref{runningexample}.
$$  
\begin{array}{l}
$$\small  
\begin{array}{l}
{\sf A}_2^{read}[user::{\bf read}(\un ,\un,\un,\un,\un)@{\sf \ds}.X] \\
\qquad \triangleq  {\bf case}( \textbf{out} \in {\sf Act}(X) )\\
\qquad \qquad {\bf break};\\
\quad \qquad{\bf proceed} 
\end{array}
$$ 
\end{array}
$$
This aspect traps a {\bf read} action of processes running at location {\sf \ds}, when reading a tuple with five fields. The process continuation of the trapped action will be recorded in variable $X$. Function \textsf{Act} returns all actions in processes represented by $X$. If the actual process bound by $X$ contains any \textbf{out} actions, the aspect will break the execution of the action and its continuation process. Otherwise, the action continues.

Suppose we have a system that contains the same net as in running example of Section \ref{runningexample} and aspect ${\sf A}_2^{read}$:

\textbf{let} $\small
\begin{array}{l}
{\sf A}_2^{read}[user::{\bf read}(\un,\un,\un,\un,\un)@{\sf \ds}.X] \\
\qquad \triangleq  {\bf case}( \textbf{out} \in {\sf Act}(X) )\\
\qquad \qquad {\bf break};\\
\quad \qquad{\bf proceed} 
\end{array}
$ \textbf{in}

$\small
\begin{array}{ll}	
& {\sf \EDB} :: \langle {\sf Alice}, {\sf MedicalRecord}, {\sf \dn}, {\sf \PA}, {\sf alicetext} \rangle  \\
\netpar
& {\sf \EDB} :: \langle {\sf Bob}, {\sf PrivateNote}, {\sf \dsn}, {\sf \RE}, {\sf bobtext} \rangle  \\
\netpar
& {\sf \ds} :: \ {\bf read}({\sf Alice}, {\sf MedicalRecord}, {\sf \dn}, {\sf \PA},!content)@{\sf \EDB}.\\
& \qquad \qquad  \ {\bf out}({\sf Alice},content)@{\sf \ds}.\\ 
& \qquad \qquad  \ {\bf out}({\sf Alice}, {\sf MedicalRecord}, {\sf \ds}, {\sf \RE},{\sf newtext})@{\sf \EDB}
\end{array}
$

and some steps of execution (omitting the aspect definition):
$$\small
\begin{array}{ll}
& {\sf \EDB} :: \langle {\sf Alice}, {\sf MedicalRecord}, {\sf \dn}, {\sf \PA}, {\sf alicetext} \rangle  \\
\netpar
& {\sf \EDB} :: \langle {\sf Bob}, {\sf PrivateNote}, {\sf \dsn}, {\sf \RE}, {\sf bobtext} \rangle  \\
\netpar
& {\sf \ds} :: \ {\bf read}({\sf Alice}, {\sf MedicalRecord}, {\sf \dn}, {\sf \PA},!content)@{\sf \EDB}.\\
& \qquad \qquad  \ {\bf out}({\sf Alice},content)@{\sf \ds}.\\ 
& \qquad \qquad  \ {\bf out}({\sf Alice}, {\sf MedicalRecord}, {\sf \ds}, {\sf \RE},{\sf newtext})@{\sf \EDB}\\

\rightarrow\\
 & {\sf \EDB} :: \langle {\sf Alice}, {\sf MedicalRecord}, {\sf \dn}, {\sf \PA}, {\sf alicetext} \rangle  \\
\netpar
& {\sf \EDB} :: \langle {\sf Bob}, {\sf PrivateNote}, {\sf \dsn}, {\sf \RE}, {\sf bobtext} \rangle  \\
 \netpar
& {\sf \ds} :: \textbf{0}
\end{array}
$$

Aspect ${\sf A}_2^{read}$ traps the {\bf read} action, whose result substitution is 
$$\small
\begin{array}{c}
[{\sf \ds}/user, \\ {\bf out}({\sf Alice},content)@{\sf \ds}.{\bf out}({\sf Alice}, {\sf MedicalRecord}, {\sf \ds}, {\sf \RE},{\sf newtext})@{\sf \EDB}/X]
\end{array}
$$

Here ${\sf Act}(P) \subseteq \{{\bf out}, {\bf in},{\bf read}, {\bf eval}, {\bf newloc}\}$ 
is the set of capabilities performed by the process $P$ (see Table \ref{PASet_functions}).
In this case,  \textsf{Act} returns set $\{\mathbf{out}\}$, the case condition evaluates to {\bf tt}, thus the aspect breaks the execution of this action and its continuation process.  		
\end{examples}


\section{Worked Examples: Advice for Usage Control} \label{sec:direct}


In this section, we now show how to use the extended AspectKE to enforce security policies that require behavior analyses of processes to be executed in the future, and we clarify the meaning of set-forming behavior analysis functions (from {\bf Set} in Table \ref{tab:syn_Aspects_process} and \ref{PASet_functions}) through examples -- enforcing several predictive access control EHR security polices to the target EHR system presented in Section \ref{runningexample}. In Section \ref{subsec:re}, we  show how to enforce policies by utilizing the set-forming functions that check properties of remotely evaluating processes. In Section \ref{subsec:pc}, we show how to enforce policies by checking properties of the continuation process at the current thread. 

\subsection{Remote Evaluation}\label{subsec:re}


Using the action {\bf eval}, AspectKE can easily express a process's remote evaluation. Moreover, using behavior analysis, AspectKE can check the content of the transmitted process by composing various aspects that embody different security considerations.  This gives us a flexible way of controlling the use of remote processes. We will enforce a security policy in a distributed mobile environment that has to consider both direct and indirect access to tuple spaces, and in the latter case AspectKE shows great usefulness.

Consider a policy concerning removal of data from the  system \cite{evered2004csa}:
\begin{qquote}
{\it Doctors, nurses and patients are not allowed to delete records from 
the database -- only the administrator of the database can do that.}
\end{qquote}
Thus, in terms of AspectKE, only the administrator is allowed to perform an {\bf in} action on the EHR database. 
\begin{examples}\label{direct_1}
In section \ref{subsec:dac}-\ref{subsec:rac}, we introduced aspects for restricting the \textbf{in} actions to access the EHR database when enforcing a basic access control policy. Here we shall slightly update them to reflect the new policy. As we prefer to use the role-based access control model, we show how to update the relevant aspects in Section \ref{subsec:rac}. 

For aspect ${\sf A}_{\sf p1_{C1}}^{read}$, we need to add role \textsf{Administrator} to role set, while updating tuple spaces \textsf{RDB} with tuple $\langle {\sf AdWalker},{\sf Administrator}\rangle$, and \textsf{PDB} with tuples $\langle {\sf Administrator, MedicalRecord},{\sf in}\rangle$ and $\langle {\sf Administrator, PrivateNote},{\sf in}\rangle$. 

In aspect ${\sf A}_{\sf p1_{C5}}^{in}$, we have forbidden all users to delete records from the EHR database. Now we relax this requirement and replace it with the following aspect. 	
$$\small
\begin{array}{l}
{\sf A}_{\sf p5}^{in}[user::{\bf in}(\un,!recordtype,\un,\un,\un)@{\sf \EDB}] \\
\qquad \triangleq  {\bf case}({\bf test}(user, {\sf Administrator})@{\sf \ER} \\
\qquad \qquad \Proceed;\\ 
\qquad \quad \Break
\end{array}
$$
This breaks direct attempts to perform {\bf in} actions, only actions by the 
administrator are allowed. 


This advice only deals with direct attempts to delete data;
we also have to cater for processes like 
$$ \small
\begin{array}{ll}
{\sf \no} :: & {\bf eval}\Big(  {\bf in}(patient,!recordtype,!author,!createdtime,!subject) @{\sf \EDB} \Big) \\
 & \qquad \quad @{\sf AdWalker}
\end{array}
$$ 
where anyone (e.g., \no) who spawns an arbitrary process on the administrator \textsf{AdWalker}'s node can behave as an administrator and delete the records.

This behavior can be captured by an aspect for {\bf eval} actions that targets the {\sf AdWalker} location, without using any behavior analysis functions.
$$\small
\begin{array}{l}
{\sf A}^{eval}_{\sf p5_A}[user::\Eval{Y}{{\sf AdWalker}}.X] \\
\qquad \triangleq 
{\bf case}({\bf test}(user, {\sf Administrator})@{\sf \ER})\\
\qquad \qquad \Proceed; \\
\qquad \quad \Break
\end{array}
$$
However, this aspect is too restrictive as it disallows the possibility for other users to perform well-behaved actions on behalf of an administrator (e.g., {\bf out},{\bf read} etc.).

Using behavior analysis functions, we are able to check the process in advance so that less restrictive policies can be enforced. For example, the following aspect    prevents remotely spawned {\bf in} actions on {\sf AdWalker}, but allows other types of action. 
$$\small
\begin{array}{l}
{\sf A}^{eval}_{\sf p5_B}[user::\Eval{Y}{{\sf AdWalker}}.X] \\
\qquad \triangleq 
{\bf case}({\bf test}(user, {\sf Administrator})@{\sf \ER})\\
\qquad \qquad \Proceed; \\
\qquad \quad  {\bf case}({\bf test}(user, \un)@{\sf \ER} \land \lnot({\bf in}\in{\sf Act}(Y)) )\\
\qquad \qquad {\bf proceed};\\  
\qquad \quad \Break
\end{array}
$$
Here ${\sf Act}(P)$ 
is the set of capabilities performed by the process $P$ (see Table \ref{PASet_functions}). It follows that any {\bf in} actions within $Y$ will be trapped. There is no restriction for actions other than {\bf in} actions, so remote code can still perform actions like {\bf out} and {\bf read}.


We may want to be more liberal and allow {\bf in} actions on 
locations distinct from {\sf \EDB}. To do so we introduce 
${\sf Loc}_{\bf in}(P)$ to be the set of locations $\ell$, where 
$\In{\cdots}{\ell}$ occur in $P$; note that this set may include location constants
as well as location variables (see Table \ref{PASet_functions}). 

Rather than aspects ${\sf A}^{eval}_{\sf p5_A}$ and ${\sf A}^{eval}_{\sf p5_B}$, we could use the aspect:
$$\small
\begin{array}{l}
{\sf A}^{eval}_{\sf p5_C}[user::\Eval{Y}{{\sf AdWalker}}.X] \\
\qquad \triangleq 
{\bf case}({\bf test}(user, {\sf Administrator})@{\sf \ER})\\
\qquad \qquad \Proceed; \\
\qquad \quad  {\bf case}({\bf test}(user, \un)@{\sf \ER} \land  ({\sf LVar_\star}\cup\{{\sf \EDB}\})\cap{\sf Loc}_{\bf in}(Y) = \emptyset )\\
\qquad \qquad {\bf proceed};\\ 
\qquad \quad \Break
\end{array}
$$
where ${\sf LVar_\star}$ is the set of all location variables in the program. This aspect allows users to perform an {\bf in} action on behalf of an administrator when the target locations will never be {\sf EHDB}, which is the least restrictive aspect for enforcing the same policy. 
\end{examples}

Example \ref{direct_1} shows that  whilst security policies  may be
 very simple to enforce superficially, execution of remotely spawned code might easily invalidate policies which appear to be reasonable.  Using aspects, we are  able to elegantly update the enforcement of the security policy to cater for this. Furthermore, the
examples also illustrate AspectKE's capability of checking the remote code before it is executed, which gives the users greater flexibility to enforce a less restrictive but more precise policy.

One might wonder whether combining the use of {\bf newloc} and {\bf eval} actions will break the above security policies. Consider the following example:
$$ \small
\begin{array}{ll}
{\sf \no} :: & {\bf newloc}(!u).{\bf out}(u,{\sf Administrator})@{\sf RDB}.\\
&{\bf eval}\Big(  {\bf in}(patient,!recordtype,!author,!createdtime,!subject) @{\sf \EDB} \Big) @ u
\end{array}
$$
Here {\sf \no} tries to create a new location, register it to {\sf RDB}, then execute the {\bf in} action from the new location. This attempt will not work: if policy ${\sf A}_{\sf p2}^{newloc}$ and ${\sf A}_{\sf p2}^{out}$ from Example \ref{example8} are still enforced, {\sf \no} is neither able to create any new location nor update {\sf RDB} since only a {\sf Manager} can do that. 


In fact, all aspects defined in this paper will directly {\bf break} any action executed at a location that has not been registered in {\sf \ER}, and this includes all attempts to use {\bf newloc} and {\bf eval} to bypass the security policies as shown above, as only the manager can update {\sf RDB} through aspects defined in Example \ref{example8}.  

%


\subsection{Using Program Continuations}\label{subsec:pc}

Now we show how AspectKE can trap the continuation process and use behavior analysis functions to get the future behavior of the executing process, which enables the advice to control and avoid executing the malicious processes as early as possible.

As we have mentioned before, our society is moving in the direction of trying to exploit 
 patient healthcare records that already exist for new purposes (known as secondary use of data). At the Canadian Institute of Health Research \cite{seconduse02} it is stated that 
``health research based on the secondary use of data contributes to our 
present level of understanding of the causes, patterns of expression and 
natural history of diseases.''
This raises new challenges for developing an effective system to ensure 
people's rights to privacy and confidentiality: decisions concerning access control 
decisions are not only based on the right of access of different principals 
but should also examine how the data is to be used after access has 
been provided.

For example, researchers who are making secondary use of data should not 
be able to access the identity of 
patients. Therefore we want to prevent them from executing a process like
$$ \small
\begin{array}{ll}
{\sf \rmi} :: & {\bf read}(patient,!recordtype,!author,!createdtime,!subject) @{\sf \EDB}  
\end{array}
$$
which explicitly specifies the $patient$ whose records the researcher is interested in reading. We can easily compose aspects that forbid the researcher from performing such actions (but allow nurses, doctors, etc
\ldots) that are similar to those aspects that have been shown in Section \ref{sec:access}.

In order for the researcher to blindly get a patient's healthcare record, the researcher
may perform a process such as
$$ \small
\begin{array}{ll}
{\sf \rmi} :: & {\bf read}(!patient,recordtype,!author,!createdtime,!subject) @{\sf \EDB}  
\end{array}
$$
The difference between the {\bf read} action in this program and those in the previous examples is the use of ! in front of $patient$, i.e., the researcher does not specify which patient's healthcare record
is to be read.
However, after the researcher has obtained the healthcare record, s/he can still \emph{use} the patient's identity. We might want to prevent the researcher from executing a process like
\begin{equation}\label{eq1}
\small
\begin{array}{ll}
{\sf \rmi} :: & {\bf read}(!patient,{\sf \MR},!author,!createdtime,!subject) @{\sf \EDB} . \\
 &  {\bf out}(patient, subject)@{\sf Publication}
\end{array}
\end{equation}
whereas it would be acceptable to execute the process
\begin{equation}\label{eq2}
\small
\begin{array}{ll}
{\sf \rmi} :: & {\bf read}(!patient,{\sf \MR},!author,!createdtime,!subject) @{\sf \EDB} . \\
 &  {\bf out}(subject)@{\sf Publication}
\end{array}
\end{equation}
since the second program will not $use$ the identity of the patient whose record has been selected.

The following policy is extensively discussed and accepted around the world and is specified directly or indirectly in a number of codes (e.g. \cite{NHSCode,seconduse02,DanishData,safran2007tnf}):
\begin{qquote}
{\it Researchers should not read and use the patient's healthcare records of an EHR system in a way that might potentially expose the identity of the patient.}
\end{qquote}
\begin{examples}\label{secondary_1}
The following aspects, which replace the aspects ${\sf A}_{\sf p1_{C1}}^{read}$  and ${\sf A}_{\sf p1_{C4}}^{read}$ in Example \ref{example5}, enforce this policy, which enforce both the basic access control policies for doctors and nurses, and policies for the researchers regarding their rights to read EHR records. It is necessary to revise the aspects because our previous development was only for primary use of data.  

$$
\begin{array}{l}

	$$\small
	\begin{array}{l}
	{\sf A}_{\sf p6_1}^{read}[user::{\bf read}(patient,recordtype,\un,\un,\un)@{\sf \EDB}.X] \\
	\qquad \triangleq  {\bf case}(
	\exists role \in \{{\sf Doctor, Nurse} \}:\\
	\qquad \qquad \qquad ({\bf test}(user, role)@{\sf \ER} \land {\bf test} (role, recordtype, {\sf read} )@{\sf \EP}))\\
	\qquad  \qquad {\bf proceed};\\
	\qquad \quad {\bf break} 
	\end{array}
	$$	
\\
	$$\small
	\begin{array}{l}
	{\sf A}_{\sf p6_2}^{read}[user::{\bf read}(patient,!recordtype,\un,\un,\un)@{\sf \EDB}.X] \\
	\qquad \triangleq  {\bf break} 
	\end{array}
	$$	

\end{array}
$$
$$
\begin{array}{l}

$$\small
\begin{array}{l}
{\sf A}^{read}_{\sf p6_3}[user::{\bf read}(!patient,recordtype,\un,\un,\un) @{\sf \EDB}.X] \\
\qquad \triangleq  {\bf case}(
\exists role \in \{{\sf Doctor, Nurse} \}:\\
\qquad \qquad \qquad ({\bf test}(user, role)@{\sf \ER} \land {\bf test} (role, recordtype, {\sf read} )@{\sf \EP}))\\
\qquad  \qquad {\bf proceed};\\
\qquad\quad
{\bf case}({\bf test}(user,{\sf Researcher})@{\sf \ER} \wedge 
\lnot (patient\in{\sf FV}(X)))\\
\qquad \qquad {\bf proceed}; \\
\qquad \quad {\Break}\end{array}
$$

\\


$$\small
\begin{array}{l}
{\sf A}^{read}_{\sf p6_4}[user::{\bf read}(!patient,!recordtype,\un,\un,\un) @{\sf \EDB}.X] \\
\qquad\triangleq
{\bf case}({\bf test}(user,{\sf Researcher})@{\sf \ER} \wedge 
\lnot (patient\in{\sf FV}(X)))\\
\qquad \qquad {\bf proceed}; \\
\qquad \quad {\Break}\end{array}
$$

\end{array}
$$

Aspect ${\sf A}_{\sf p1_{C1}}^{read}$ is replaced/divided by aspects ${\sf A}^{read}_{\sf p6_1}$ and ${\sf A}^{read}_{\sf p6_3}$, according to whether a patient name is clearly specified or not. Similarly, ${\sf A}_{\sf p1_{C4}}^{read}$ is replaced/divided by aspects ${\sf A}^{read}_{\sf p6_2}$ and ${\sf A}^{read}_{\sf p6_4}$. In both cases, additional conditions regarding researchers are only added to aspects when a patient name is not specified, i.e., ${\sf A}^{read}_{\sf p6_3}$ and ${\sf A}^{read}_{\sf p6_4}$. In these two aspects, the behavior analysis function \textsf{FV} is used, which returns the set of free variables of $P$  (see Table \ref{PASet_functions}). 

In our case the aspect ${\sf A}^{read}_{\sf p6_3}$ will bind $X$ with the {\bf out} actions of the above two programs: in program (\ref{eq1}) with ${\bf out}(patient,subject)@{\sf Publication}$, and in program (\ref{eq2}) with ${\bf out}(subject)@{\sf Publication}$. Thus $\lnot (patient\in{\sf FV}(X))$ would be evaluated to $\FF$ for the first case and to $\TT$ for the second case. Using this extra information from the behavior analysis function, the advice can be based on some properties of the future execution of the continuation process (in particular, whether or not $patient$ will ever be used). And the suggestions from aspect ${\sf A}^{read}_{\sf p6_3}$ would be {\bf break} for the first case and {\bf proceed} for the second one. Note at the point that the access control decision has been made, $!patient$ in the join point {\bf read} action is still a defining occurrence variable and thus does not bind with any actual location yet, and behavior analysis merely analyses the future behavior of process based on its static information.
\end{examples} 


The above aspect is too restrictive since it forbids  the
execution of meaningful {\bf read} actions as well. As one of the case studies performed by 
the Canadian Government \cite{seconduse02} indicates:
\begin{qquote}
{\it In practice, the researchers might need to do some data linkage operations between different databases.}  
\end{qquote}
For example,  we may allow the researcher to extract several records for the same patient from different databases and put that information together as in the process
\begin{equation}\label{eq3}
\small
\begin{array}{ll}
{\sf \rmi} :: & {\bf read}(!patient,{\sf \MR},!author,!createdtime,!{\it subject1}) @{\sf \EDB}.  \\
&  {\bf read}(patient,{\sf \MR},!author,!createdtime,!{\it subject2}) @{\sf \EDBT}.  \\
 &  {\bf out}({\it subject1}, {\it subject2})@{\sf Publication}
\end{array}
\end{equation}
where we introduce another EHR database {\sf \EDBT} that is located at another hospital.

Now there are two databases so we need to consider which policies are suitable for each of them, respectively. For illustration purposes, we might simply demand that the second database has the same security policy as the original one. Thus the second {\bf read} action would be denied, since the original policy prohibits  a researcher from  reading a specific patient's healthcare record.  Because of this, establishing data linkage in the direct manner of program (\ref{eq3}) will never work.

{ 
To make the data linkage work we can restrict the researcher's access to the data through a trusted location ({\sf \EDB} for example) by remote evaluation, and let the trusted location perform the actual data linkage actions. In this way the policy will allow the second {\bf read} action whenever it is executed from the trusted location: 
\begin{equation} \label{eq4} \small
\begin{array}{ll}
{\sf \rmi} :: & {\bf eval}\Big({\bf read}(!patient,{\sf \MR},!author,!createdtime,!{\it subject1}) @{\sf \EDB}.  \\
& \qquad {\bf read}(patient,{\sf \MR},!author,!createdtime,!{\it subject2}) @{\sf \EDBT}.  \\
 & \qquad  {\bf out}({\it subject1}, {\it subject2})@{\sf Publication}\\
& \qquad \Big)@{\sf \EDB}
\end{array}
\end{equation}}

As demonstrated in the last subsection, if remote evaluation is allowed, we need to pay closer attention to the overall security of the system.  For example, in the event that the researcher attempts to get a doctor to link databases and output the private information as in
\begin{equation} \label{eq5} \small
\begin{array}{ll}
{\sf \rmi} :: & {\bf eval}\Big({\bf read}(!patient,{\sf \MR},!author,!createdtime,!{\it subject1}) @{\sf \EDB}.  \\
& \qquad {\bf read}(patient,{\sf \MR},!author,!createdtime,!{\it subject2}) @{\sf \EDBT}.  \\
 & \qquad  {\bf out}(patient, {\it subject1}, {\it subject2})@{\sf Publication}\\
& \qquad \Big)@{\sf \ds}
\end{array}
\end{equation}
or in the event that the researcher attempts to obtain the records of a patient whose name is selected from his own tuple space either before evaluating a process at an EHR database or during the evaluation procedure.
\begin{equation} \label{eq51} \small
\begin{array}{ll}
	{\sf \rmi} :: & {\bf read}(!patient)@{\sf \rmi}.  \\	
	&  {\bf eval}\Big({\bf read}(patient,{\sf \MR},!author,!createdtime,!{\it subject1}) @{\sf \EDB}.  \\
	 & \qquad  {\bf read}(patient,{\sf \MR},!author,!createdtime,!{\it subject2}) @{\sf \EDBT}. \\
	 & \qquad  {\bf out}(patient, {\it subject1},{\it subject2})@{\sf Publication}\\
	& \qquad \Big)@{\sf \ds}	
\end{array}
\end{equation}
\begin{equation} \label{eq52} \small
\begin{array}{ll}
	{\sf \rmi} :: & {\bf eval}\Big( {\bf read}(!patient)@{\sf \rmi}.  \\	
	&  \qquad {\bf read}(patient,{\sf \MR},!author,!createdtime,!{\it subject1}) @{\sf \EDB}.  \\
	 & \qquad  {\bf read}(patient,{\sf \MR},!author,!createdtime,!{\it subject2}) @{\sf \EDBT}. \\
	 & \qquad  {\bf out}(patient, {\it subject1},{\it subject2})@{\sf Publication}\\
	& \qquad \Big)@{\sf \ds}	
\end{array}
\end{equation}
we need an aspect that inspects the actions performed by a researcher performing an {\bf eval} action on all the EHR databases or locations other than the EHR databases.
\begin{examples}\label{secondary_3}
This motivates the aspect:
$$\small
\begin{array}{l}
{\sf A}^{eval}_{\sf p7}[user::\Eval{Y}{dest}.X] \\
\qquad \triangleq  {\bf case}({\bf test}(user, {\sf Researcher})@{\sf \ER} \land {\bf test}(dest, {\sf DataBase})@{\sf \ER}) \wedge \\
\qquad \qquad \qquad  \forall x \in {\sf LC}_{\bf read}(Y) :  \lnot({\bf test}(x, {\sf Patient})@{\sf \ER}) \wedge\\
\qquad \qquad \qquad  \forall y \in {\sf Loc}_{\bf read}(Y) :  ({\bf test}(y, {\sf DataBase})@{\sf \ER}) \ )\\
\qquad \qquad \Proceed; \\
\qquad\quad
{\bf case}({\bf test}(user, {\sf Researcher})@{\sf \ER}  \wedge \\
\qquad \qquad \qquad  (\lnot {\bf test}(dest, {\sf DataBase})@{\sf \ER} \land  {\bf test}(dest, {\un})@{\sf \ER}  )  \wedge \\
\qquad \qquad \qquad  \forall x \in {\sf Loc}_{\bf out}(Y) :  ({\bf test}(x, {\sf DataBase}) \lor x \in \{ dest \} )\ )\\
\qquad \qquad \Proceed; \\
\qquad \quad  {\bf case}(\lnot {\bf test}(user, {\sf Researcher})@{\sf \ER} \land {\bf test}(user, \un))\\
\qquad \qquad \Proceed; \\
 \qquad  \quad \Break
\end{array}
$$
{
The first case of the aspect ensures that a researcher can directly evaluate processes in the EHR databases if s/he is not able to get the name of the patient whose record s/he is trying to obtain either before evaluation or during evaluation (by reading a patient name from a location other than EHR databases). Note that  {\sf LC}$_{\bf read}(P)$ is a set of location constants in read actions of process P (defined in Table \ref{PASet_functions}).  The second case guarantees that when sending a process to other remote locations, the process only contains {\bf out} actions that are performed on the EHR databases (trusted locations) or the tuple space associated with that remote location. 
}
\end{examples}

In summary, the examples in this section show the versatility of AspectKE when dealing with remote
evaluation and future execution paths of processes. We illustrated the usefulness of each behavior analysis function when enforcing access control policies that requires the future behavior of process. When selecting the appropriate behavior analysis functions, it is possible to enforce less restrictive policies and avoid the execution of malicious processes as early as possible.    
More importantly, in a highly complex and privacy related computing environment, with policies that are changed frequently such as the EHR system, enforcing various access control and data usage policies through security aspects shows the potential of being a very flexible and elegant approach.


\section{Related Work} \label{sec:related}

\subsection{Policy Enforcement Mechanisms and Aspect Oriented Programming}
There are various techniques for enforcing security policies, the most  traditional one is  a \emph{reference monitor} that observes software execution and dynamically mediates all access to objects by subjects. Instead of mixing the monitoring code in the target system, Inlined Reference Monitors (IRMs) \cite{erlingsson2000iej} use a  load-time, trusted program rewriter to insert security code into a target  application, resulting in a self-monitoring application that performs security checks as it executes. There are many IRM systems implemented by various program rewriters (e.g. \cite{Ulfar99,evans1999fpd,bauer2005csp,hamlen2006spe}),  ensuring that different types of application will obey their corresponding security policies.  

Independently, the aspect-oriented programming (AOP) paradigm \cite{kiczales2001oa} has emerged and has  served as another effective mechanism for tackling the same issue. Indeed security is naturally  identified as one kind of  \emph{crosscutting concern} that aspect-oriented programming was designed to deal with: instead of using a rewriter to inject monitoring code, security policies are coded through aspects which are invoked  when the target program executes certain actions. There is a close connection between AOP and IRM. Hamlen and Jones \cite{hamlen2008aol} propose an aspect-oriented security policy specification language SPoX for enforcement by IRMs which establish a formal connection between AOP and IRMs. JavaMOP\cite{Chen05javamop} implements IRM by using AspectJ aspects as the instrumentation mechanism. AspectKE takes the AOP approach to internalize the reference monitor for enforcing security policy to tuple space systems, and directly encodes security concerns in aspects. 

Most research focuses on the class of security policies that can be enforced by monitoring execution of a target system \cite{schneider2000esp} and hence are enforceable by traditional reference monitors.  
AspectKE allows us to perform a behavior analysis on future execution of the target system, giving us the capability  of enforcing policies that go beyond reference monitors. In \cite{schneider2000lba}, the authors outline several promising methods for enforcing security policies: IRM, type systems and certifying compilers. The authors also argue that synergies among these approaches will achieve remarkable results. We believe our approach --  aspects with behavior analysis -- is comparable as an alternative to IRM with Type systems.

Several AOP languages can identify the data-flow and control-flow between join points, which can serve as powerful policy enforcement mechanisms. AspectJ's cflow\cite{kiczales2001oa} captures the control flow between join points. Dataflow pointcut \cite{masuhara2003dpa} identifies join points based on the dataflow of information. Tracematches\cite{allan2005adding} can give advice based on the execution history of computation. However, these systems can only refer to the past and current events, in contrast AspectKE can refer to future events. A few AOP languages propose mechanisms so that aspects can be triggered by  control flow of a program in the future, e.g, \texttt{pcflow}\cite{kiczales2003fun} and \texttt{transcut}\cite{sadat2009transactional}, however, they lack support for providing dataflow information in the future as AspectKE does.  Some advanced AOP languages (e.g.,\cite{Aotani07, chiba04}) offer ways of referring to the future behavior of a program in the aspect, which in theory can be used for specifying security policies that depend on the future control-flow and data-flow. However, as they usually lack formal semantics and also normally only offer access to low level (e.g., bytecode-level) information of a program, this makes it hard to understand and develop appropriate underlying analyses for enforcing security policies. Moreover, they lack high level abstraction for presenting the analysis results as AspectKE's behavior analysis functions do, which makes it non-trivial to use the results for composing security policies. The formal semantics of AspectKE clarifies the way of developing useful behavior (program) analyses and presenting the analysis results through appropriate language abstraction, and formally pave a way of integrating program analysis techniques into the policy specification and enforcement procedure.

\subsection{Security With Aspect-Oriented Programming}
There are many papers that explore the AOP mechanism to enforce security policies. 
One line of work directly or indirectly use the popular Java-based general purpose AOP languages like AspectJ \cite{kiczales2001oa}, Hyper/J \cite{ossher2000hjm},  CaesarJ \cite{aracic2006oc} to express and enforce security policies. 

In \cite{dewin2002sta} the authors present general guidelines for how to compose access control aspects in AspectJ, while in \cite{Tine05} an enforcement of application-specific policies in an access control service is implemented in CaesarJ. 
Phung and Sands \cite{phung:spe} identify classes of reference monitor-style policies that can be defined and enforced by AspectJ and present a method to realize some history-dependent security policies which cannot be naturally expressed in AspectJ. 
Ramachandran et al. \cite{ramachandran2006ams} discuss using AspectJ for implementing multilevel security and demonstrate how aspects, in comparison to traditional programming, can guarantee better security assurance. Similarly, AspectKE can be used to enforce a wide range of security policies but focuses on access control and explicit flow of information.

Oliveira et al. \cite{deoliveira2007wrb} use their own rewrite-based system to express access control policies and then map them into an AspectJ program; in \cite{cuppens2006aeo}, availability requirements are  expressed in a formal model that combines deontic and temporal logics, and are then translated into availability aspects in AspectJ. One advantage of these approaches is that policies can be formalized through security oriented languages which are more suitable for security considerations than general purpose languages, another advantage is that some policy languages have a formal semantics which enables formal verification. AspectKE is designed with security in mind and has a formal semantics which will enable us to reason about the AspectKE's policies in future work.

Even though these well-known AOP languages are industrial strength and can be readily used for a policy enforcement mechanism,  they have their limitations:  it is very hard to apply these AOP languages to the types of systems we have been studying. For example, the languages are designed for programs that run on a local machine, and they do not naturally support pointcuts for a distributed system. They also lack a pointcut mechanism to capture the future execution of a program, and thus are unable to enforce predictive access control policies. These issues are explicitly addressed by AspectKE. Particularly, we are the first to report how to use aspect oriented programming techniques to enforce secondary use of data policies that are becoming increasingly important in large IT systems. 


As we have done, some researchers design their own special purpose aspect languages or systems to study security enforcement mechanisms. 
For example, HarmlessAML \cite {Daniel07} is an  aspect-oriented  extension of Standard ML, and has a type system that guarantees well-typed harmless advice does not interfere with mainline logic computation. AspectKE is an aspect-oriented extension of a tuple space system,  with a different research focus, aim at enforcing access control policies to a distributed computing model and studying how properties obtained from the behavior analyses can be used to specify access control policies.  
As mentioned earlier, in \cite{masuhara2003dpa} the dataflow pointcut is proposed which specifies where aspects should be applied based on the origins of values in the past execution, which is useful in a situation where flow of information is important. AspectKE tackles similar issues, but checks the flow of information in the future.

\subsection{Distribution with Aspect-Oriented Programming}
Much work has been done regarding how to deploy and weave aspects for distributed systems: some work is relevant for language design of  distributed AOP with explicit distribution \cite{nishizawa2004rpl,BENAVIDESNAVARRO:2006:INRIA-00071386:1,tanter2006versatile}, other work explores the implementation of AOP middleware to support distributed AOP \cite{pawlak2004jac,lagaisse2006true,Eddy2008}. AspectKE is closer to the language design of distributed AOP which naturally follows the KLAIM programming model and uses remote pointcuts \cite{nishizawa2004rpl} that identify join points in a program running on a different location. However, AspectKE does not aim at enhancing the flexibility of mechanisms to deploy, instantiate and execute distributed aspects, e.g.,  support advice execution over remote hosts, as AWED \cite{BENAVIDESNAVARRO:2006:INRIA-00071386:1} and ReflexD \cite{tanter2006versatile} have achieved, rather it focuses on integrating analysis components for reasoning about the local or mobile processes to support advanced access control in a distributed setting. Compared with these languages, AspectKE provides a well defined security enforcement mechanism to tuple space systems that supports process mobility.

AO4BPEL\cite{charfi2004aow} is an aspect extension of the process-oriented composition language BPEL, which was originally designed for composing Web Services. 
Work in \cite{capizzi2004eec,colman2005csr} discusses different principles of  using AOP to implement coordination systems (in AspectJ), but that are not related to security. 
Recently, another variant of AspectK, AspectKB \cite{chris09} has been proposed; it uses Belnap Logic to deal with conflicts when distributed advices are composed in a coordination environment.   

\subsection{Security in Coordination Languages and Security Policy Languages}

AspectKE extends KLAIM, first presented in \cite{De1998}, later evolved into the KLAIM family (reviewed in \cite{bettini2003kpt}), including cKlaim, OpenKlaim, HotKlaim, OKlaim and X-Klaim etc. The prototype language of KLAIM is Klava \cite{bettini2002kjp}, implemented in Java and has proved to be suitable for programming many distributed applications involving code mobility and our AspectKE* prototype language \cite{fan10} is built on top of it. Regarding policy enforcement of these languages, some authors use control and data flow analyses that are written in the Flow Logic approach (e.g.~\cite{hansen2006sm,HansenNNP08}),  others use type systems (e.g.~\cite{denicola2000tac}),  and  \cite{rocco08} combine these two lines of work.  They can be used to enforce very advanced security policies, however,  all of them require the user to explicitly annotate policies in the main code  (e.g. attach policies to each location), while our approach avoids this by specifying them inside the aspects, thus achieving a better separation of concerns. 

Secure shared date-space coordination languages can be classified into two categories with regard to the underlying access control mechanisms \cite{focardi2006secure}: the entity-driven approach (additional information, associated to resources such as tuple spaces, tuples and single data fields, list the entities which are allowed to access the resources) e.g., Secure Lime \cite{handorean2003secure} and KLAIM \cite{De1998};  and the knowledge-driven approach (resources are decorated with additional information and the processes can access the resources only in the event that they prove to keep the knowledge of such additional information) e.g., SecOS \cite{VitekBO03} and SecSpaces \cite{gorrieri2006supporting}. Our approach is suitable for expressing access control policies that fit both an entity-driven approach and a knowledge-driven approach, as the additional information is essentially expressed in aspects and is directly embedded in neither  resources nor processes. Moreover, this additional information is not limited to the past and current facts used in the previous work, e.g.,  password \cite{handorean2003secure}, locks \cite{VitekBO03} or partitions \cite{gorrieri2006supporting}, but also facts about the future, e.g., how particular data will be used, which is useful for enforcing predictive access control policies that only AspectKE can enforce.    

Binder \cite{detreville2002blb} and Cassandra \cite{becker2004cft} are   very powerful logic-based security policy languages, which are both based  on the datalog logic-programming language. In \cite{becker2004cft}, there is a  substantial case study performed using Cassandra that is based on  the UK National Health Service procurement exercise.  In AspectKE,  our security policies are mainly expressed based on logical formulae and predicates in the aspects. The difference is AspectKE also provides predicates that can also describe future  events.  There are other prominent policy languages like Protune \cite{bonatti14dam},  Rei \cite{kagal2003pba}, Ponder \cite{damianou2000lss}, and  KeyNote \cite{blaze1999rkt}, which can express basic access control policies  very well.  Only Ponder and Rei can express usage control through  obligation policies but, unlike our approach,  neither language  can enforce them and has to trust that the party receiving the data uses it in proper ways \cite{duma2007psw}.




\section{Conclusion} \label{sec:conclusion}

\subsection{Our Experience of a Proof-of-Concept Implementation}

We have designed and implemented a proof-of-concept programming language AspectKE* \cite{fan10}, based on the core concepts of AspectKE, which can be compiled and executed under a Java environment for building secure distributed systems and is freely distributed\footnote{http://www.graco.c.u-tokyo.ac.jp/ppp/projects/aspectklava.en}. The runtime system of AspectKE* is built on top of Klava \cite{bettini2002kjp}. Klava, a Java package that implements the core concept of KLAIM, can be used to program tuple space operations for building distributed systems. 

AspectKE* is generally implemented following the semantics of AspectKE, however, it takes a different approach regarding the implementation strategy of the behavior analyses. When aspects need the results of behavior analysis functions, AspectKE needs to perform the corresponding behavior analyses at runtime for each step of program execution, which is not practical enough due to the potential large runtime overhead. On the other hand, AspectKE* uses a two-staged implementation strategy that gathers fundamental static analysis information at the process's load-time (the analyses are performed on the bytecode representation of the processes),  and evaluates the program analysis predicates and functions (similar to the behavior analysis functions) at runtime by combining the results of the load-time analysis and runtime information, which efficiently implements the conceptual model of AspectKE.

We implemented a secure tuple space based EHR workflow system in AspectKE*, and all aspects presented in the previous sections are implemented, except for aspect ${\sf A}^{eval}_{\sf p7}$, as the behavior analysis functions {\sf LC} and {\sf LC$_c$} are not currently supported by AspectKE*. Besides the EHR system, we also built a secure tuple space based chat application in AspectKE*. In \cite{fan10} we show how to enforce access control policies  which require analysis of future behavior of a process  onto a chat system that contains untrusted components. These two applications demonstrate that the AspectKE model can be efficiently implemented and executed in real world settings. 

\subsection{Final Words and Future work}
We have presented AspectKE, an aspect-oriented extension of the coordination language KLAIM \cite{bettini1998ima}. 
This has provided a concrete vehicle for presenting our approach; the distributed
tuple spaces provide a natural model of the kind of system that motivated
our work.  However, the approach could equally well be applied to more
classical process algebraic languages; the join points in this case being
read and write accesses to channels.

Compared with our previous work AspectK \cite{chris08}, AspectKE 
empowers aspects to trap not only the matched actions but also the process continuation and the processes being evaluated remotely. AspectKE also provides various behavior analysis functions and enables us to reason about the future execution of processes, which improves the capability of standard reference monitors that normally only deal with history based security policies. To achieve this, we simplify AspectK so that actions before and after {\bf proceed} or {\bf break} are not allowed. If we had allowed these actions, a safe behavior analysis would be very difficult to achieve, since the  processes \emph{to be executed} might execute more actions (inserted by aspects at runtime) than planned.  This is an interesting direction for the future work and will require more powerful program analyses than the behavior analyses of the present paper.  

We evaluated the expressiveness of AspectKE by investigating its policy enforcement capability through examples in an EHR setting. We have demonstrated that AspectKE can enforce discretionary access control, mandatory access control as well as role-based access control. Furthermore, AspectKE can elegantly retrofit new policies to an existing system with minimum effort at any phase during the system development cycle. We have also shown that in a distributed and mobile system,  the information flow is very hard to control. AspectKE can enforce a range of predictive access control policies to cater for this issue, through composing aspects that check remote evaluation and the program continuation.  We enforce  both primary and secondary use of data, which shows that AspectKE is suitable to cater for old as well as new challenges in such a complex distributed computing environment, where security and privacy are of great importance. We also briefly report the experience of our proof-of-concept implementation of AspectKE, namely AspectKE*, and its usage for enforcing the security policies presented in this paper to an EHR workflow system and enforcing other security policies to a chat application.       

The examples presented in this paper were chosen so as to illustrate the different characteristics of AspectKE, but they were also chosen so as to constitute a complete set of access control policies. Here we conjecture that our aspects indeed enforce the complete set of access control policies, but in order to validate our conjecture we need to apply a formal validation method to it.   

There are  other challenging secondary use of data policies which require control not only of \emph{direct flows} but also of \emph{indirect flows} 
\cite{sabelfeld2003lbi}: e.g. after storing specific data into a doctor's 
own tuple space,  the doctor should not allow them to be transferred into a researcher's 
tuple spaces by indirectly passing through another location.  In this case, checking all the parallel executing processes with existing security aspects  that rely on more advanced static analyses  \cite{NNH_PPA} might be needed (e.g. to predict all possible data that can be stored in a certain tuple space \cite{hansen2006sm,HansenNNP08}).  

We find that the combination of aspects with 
behavior and static analysis techniques shows great potential for serving as a 
flexible and powerful mechanism for policy enforcement, and as a 
promising method of building security and trust in a distributed and mobile environment.

%
%




\section*{Acknowledgement}
This project was partially funded by the Danish Strategic Research Council (project
2106-06-0028) ``Aspects of Security for Citizens''. We would like to thank Hidehiko Masuhara and Tomoyuki Aotani for working with us in AspectKE*, and for their comments on the paper.







\bibliographystyle{elsarticle-num}
\bibliography{refs}



\end{document}